# A Survey of Self-Sovereign Identity Ecosystem

**Reza Soltani, Uyen Trang Nguyen, and Aijun An**

*Lassonde School of Engineering, Department of Electrical Engineering and Computer Science, York University, Toronto, Canada*

Correspondence should be addressed to Reza Soltani; rts@cse.yorku.ca

Self-sovereign identity is the next evolution of identity management models. This survey takes a journey through the origin of identity, defining digital identity and progressive iterations of digital identity models leading up to self-sovereign identity. It then states the relevant research initiatives, platforms, projects, and regulatory frameworks, as well as the building blocks including decentralized identifiers, verifiable credentials, distributed ledger, and various privacy engineering protocols. Finally, the survey provides an overview of the key challenges and research opportunities around self-sovereign identity.

## 1. Introduction

Existing identity management models are not future proof. The existing digital identity management models which rely on centralized data repositories and identity providers have led to a growing number of data breaches, causing significant loss of personal data and tremendous cost to all stakeholders and in particular the users. The proliferation of digital services has placed digital identity in the center stage. As the use of online services increases, we are faced with the growth of a complex, inconsistent, tangled, and insecure web of digital identity practices. As the number of identity breaches grows, we are observing the increased awareness in implications associated with the existing digital management approaches and their deficiencies. This has led to seeking more secure and privacy-preserving approaches to digital identity management. The idea behind self-sovereign identity is the ability for the identity holders (e.g., users) to have better control over their identity data, with strong emphasis on data portability and data minimization. Lack of adequate data ownership and control over digital identity data by users, in addition to the absence of adequate digital identity for over a billion individuals worldwide [1], have a major negative impact on users' privacy rights and access to services [2]. Moreover, password-based authentication methods continue to be one of the most common approaches of user authentication to online services, leading to an array of usability and security impediments. In addition, the retention of users' identity data on multiple centralized repositories, with diverse implementations of security and privacy standards and lack of adequate standard in data management practices, form enticing targets for hackers to exploit, leading to growing security breaches and identity fraud.

Lastly, in jurisdiction with inadequate privacy laws and protection, data repositories are subject to government's arbitrary access to identity data and extrajudicial surveillance without the prior consent of the user. When users share their identity data to an organization, they lose visibility and control over how their data are stored or accessed. Depending on an organization's jurisdiction of operation and data stores, they may be subject to federal legislation that will require them to provide access to their customer data, without the consent of the customer. This concern led the Court of Justice of European Union in July 2021 to determine in the case of Data Protection Commission v. Facebook Ireland Ltd. and Maximillian Schrems that the EU-U.S. Privacy Shield Framework for data transfers is not adequate. The reason being that the nature of the U.S domestic law gives the government and authorities arbitrary power to access private sector data. The court found that the U.S. surveillance programs allow the government access to personal data are not limited to what is strictly necessary and proportional as requested by EU law [3].

While the cost of identity and access management (IAM) is increasing each year, surpassing $3.3 billion [4], users are

also becoming more aware of the value of their data, and the importance of adequate privacy control. Recent developments in identity protocols, technologies, and specifications have excited many in the industry and academia to believe the technical foundation for the next breakthrough in digital identity management is in our grasp.

This article provides a comprehensive survey of the research landscape for self-sovereign identity and delivers an overarching overview of the technologies, standards, industry initiatives, and academic work related to this rapidly evolving topic. To this end, the article starts by offering the definition of the digital identity and access management system, iterating through various IAM models leading up to the SSI model. The article then provides the building blocks of the SSI model, in addition to the current technologies and academic literature work. With this contribution, we aim to provide a comprehensive survey of the SSI ecosystem, clarify its definition and distill away the unnecessary noises surrounding it. Lastly, we aim to objectively highlight the importance and benefits of SSI, as well as its shortcomings, in order to promote more academic work on this topic. While SSI is a major stepping stone towards a more user-centric and privacy-respecting design thinking, it is still in its infancy. The SSI model is not the panacea to every problem in the digital identity and privacy space, nor does it guarantee a complete solution to every open problem presented in this survey. We should also emphasis the lively, rapidly expanding quality of the SSI ecosystem, with new prototypes, conferences, and ideas emerging frequently.

While we strive to provide a comprehensive and accurate representation of the SSI space, we refer the reader to resources referenced in this article to obtain the very latest on its development.

The paper is organized as follows: in Section 2, we present our research motivation by stating the driving factors witnessed in the current digital identity climate. In Section 3, we provide an in-depth definition of identity and digital identity as well as the architecture of the identity and access management model. This is followed by a summary of various iterations of digital identity management models starting with the isolated identity model and ending with the emerging self-sovereign identity model. The section concludes by discussing the principles and challenges of the SSI model. In Section 4, we provide an architectural overview of the SSI model and its building blocks and concepts, including decentralized identifiers, verifiable credentials, and digital wallets. The section also describes blockchain technology and its role in realizing the SSI requirements. In Section 5, we present a summary of various privacy engineering techniques complementing or extending the SSI model. Among these techniques are Zero-Knowledge Proofs protocol and Threshold Secret Sharing schemes. In Section 6, we present a list of known platforms and technical groups in the SSI ecosystem that significantly contribute to the growth and maturity of this model. In Section 7, we discuss emerging regulations, laws, and policies relevant to SSI and, in particular, the laws emerged from the European Union and the United States. Section 8 provides a technical summary of related research surveys and prototypes in the context of SSI. In Section 9, we state our future research directions and concluding remarks. Finally, Section 10 provides the appendix of common terms mentioned in this article.

## 2. Challenges of Current Digital Identity Management Models

In this section, we introduce a number of challenges found in current identity management systems that contribute to driving the SSI adoption forward. They include inadequate data ownership and control by *data subjects*, the limitations with using password-based authentication, fragmentation of data, and finally the growing rate of data breaches and fraud. Data subject is an identifiable natural person who can be identified, directly or indirectly, in particular by reference to an identifier such as a username or an identification number [5]. In this paper, we use the term "data subject" and "user" to refer to the entity interacting with online service providers.

*2.1. Data Ownership and Governance.* Based on the World Bank estimates, there are over 1.1 billion individuals who lack an official identity [1] and 3.5 billion people worldwide who are underbanked or unbanked [6]. For individuals who do have online accounts, the existing identity models do not place them in immediate control of their identity data, as those individuals have limited control or visibility into how their personal data are managed, shared, and discarded by service providers. As Shoshanna Zuboff describes in her book The Age of Surveillance Capitalism [7], many identity providers generate income by collecting behavioral data of their users, which they use to develop sophisticated user behavior analysis and prediction systems. These are then traded on marketplaces where advertisers can select a fitting target audience for their products. Presently, the majority of online service providers support their own specific set of data management policies and practices, leading to industry monopolies and vendor lock-ins. As a side effect, this can cause some service providers to deny access to the data subjects at any time, for reasons they deem justified. This has led to many cases of account closures and deactivation by the service providers with little explanation to the data subjects [8]. These drawbacks have led the pursuit of new identity models, processes, and regulations with the goal of providing more control and choice to the data subjects, and more transparency over methods by which service providers handle personal data. While certain revocation and termination of user accounts can be considered acceptable, unauthorized analysis and sharing of personal data by service providers can be avoided if data subjects have the option of maintaining their personal data on their own trusted devices or have the ability to delegate this task to a data custodian of their choosing that conforms to the same open standards as the rest of the network.

*2.2. Password-Based Authentication.* While many data subjects are familiar with the possibility of identity fraud and its ramifications, they continue to choose weak passwords for their online accounts. This is partly due to the overwhelmingly large number of online services that need to be managed by a subject. On average, a LastPass business user has to track 191 passwords [9]. Due to the difficulties associated with account management, many data subjects leverage their social accounts such as Facebook and Google to authenticate with online services, making their social media accounts prone to more risk than before. In most cases, a password-based authentication system is easier to set up than a two-factor authentication system. Until alternative methods to using passwords as credentials, such as biometrics, become as easy to deploy and use, the use of passwords will be the primary method of authenticating data subjects.

*2.3. Fragmented Identity Data.* The proliferation of information in the digital economy has led to the fragmentation of identity data among a myriad of online identity repositories. An average data subject has their data scatters among various governmental, financial, and social data hubs. Duplicate entries, mismatches, and outdated data add to the difficulty to maintain all identity repositories. Often a single organization is fragmented by business units or products, which adds to this complexity. The privacy and security due diligence of an identity repository depends on its business strategy, goals, and budgeting plans. Moreover, lack of universal standards and interoperability among the identity repositories makes it challenging for the data subjects to store, obtain, remove, or share their personal data.

*2.4. Data Breaches and Identity Fraud.* Nowadays, data breaches have become part of our daily lives. In the first half of 2018 alone, there were 944 breach incidents reported, affecting more than 3.3 billion records [10]. Based on the Javelin Strategy and Research 2018 report [11] in 2017, more than 16 million U.S. consumers were affected by identity fraud which resulted in $16.8 billion in damages. In 2017, the Equifax incident alone affected more than 146 million users [12]. In 2018, more than 2.1 billion Facebook records were potentially compromised, and 336 million Twitter credentials were exposed in plain-text. Moreover, the outdated methods in dealing with physical identity documents introduce their own unique privacy and security issues. Physical documents can be falsified, altered, lost, or stolen, and their presentation and transfer can lead to human errors.

The above factors motivate us to embark on a research voyage to explore emerging digital identity and access management concepts and examine the rapidly emerging self-sovereign identity management model as a potential solution to addressing some of the aforementioned shortcomings.

In the next section, we describe the fundamentals of digital identity and the overall architecture of a digital identity system.

## 3. Identity and Access Management

In this section, we begin by describing the concept of identity, digital identity, and digital identity and access management systems. We provide a description of IAM architecture and its key operations, and a detailed explanation of existing identity management models leading up to self-sovereign identity.

*3.1. Identity.* Identity has a long story. Possessing an identity has been of importance from the early times [13]. Depending on the context, the term identity is defined differently [14]. The American Heritage Dictionary defines identity as “The set of characteristics by which a person or thing is definitively recognizable or known,” and “The fact or condition of being the same as something else.” [15].

The word identity originated from the Middle French “identité”, which is derived from Late Latin “identitat-, identitas,” from Latin “identidem,” which is a contraction of “idem et idem,” meaning “same and same” [16].

Although the phrase “identity” is commonly used in everyday discourse and various context, it is proven rather difficult to provide an adequate and concise explanation to encompass its every meaning [17].

The French philosopher Paul Ricoeur defines identity as two-fold. The first is known as *Idem* which means sameness in Latin [18], and the second aspect is defined as *Ipse* meaning selfhood in Latin [18]. The sameness feature represents the characteristics that remain unchanged, and the selfhood aspect represents the features that are unique. From that definition, identity can be defined as the representation of an *entity* through features that make the entity unique and/or persist through time, in a given context. The term entity is defined as a logical or physical object with a separate distinctive existence either in a logical or physical sense [19].

In psychology, the term is generally referred to all the psychological traits of a person, including beliefs, personality, and other attributes [20]. While in sociology, the concept of identity includes the history, culture, religion, and traditions associated to the person [21].

The modern definition of identity as we know it was first used in 1950 by Erik Erikson and has been used in various contexts and purposes [22]. Glässer and Vajihollahi define identity as a logical representation of a physical presence of a person or thing [23]. Wang and Filippi define identity as all attributes of a person that uniquely defines the person over the course of a lifetime, providing sameness and continuity despite varying aspects and conditions [24].

In the book “Identity is the New Money” [25], David Birch projected the next evolution of identity to be defined by our social graphs, which take the model back to a digitized version of the preindustrial approach. The social graph is a digital representation of our truthful relationships.

*3.2. Digital Identity.* A digital identity is the unique representation of an entity within a particular digital context.

The National Institute of Standards and Technology (NIST) [26] defines digital identity as “the unique

representation of a subject engaged in an online transaction. A digital identity is always unique in the context of a digital service but does not necessarily need to uniquely identify the subject in all contexts. In other words, accessing a digital service may not mean that the subject's real-life identity is known." The International Telecommunication Union (ITU) [27] defines digital identity as "A digital representation of the information known about a specific individual, group, or organization."

Cameron defines digital identity as "a set of claims made by one digital subject about itself or another digital subject" [28]. He also introduces the *Seven Laws of Identity*. These laws are as follows:

(1) Users should be in control of how their identity information is shared

(2) The amount of information disclosed should only be the minimum necessary amount required, and data should not be kept longer than needed by the other entities

(3) The user should be well informed about which entities manage their identity information

(4) The user's information should not be created or exposed in such a way to allow data correlation, pattern recognition, or entity identification by unauthorized entities

(5) Interoperability and seamless integration among various entities supported by different architecture should be possible

(6) Reliable and secure integration between human users and machines should be empowered

(7) Consistent user experience across multiple contexts and technologies

A *partial identity* or *persona* is a subset of identity attributes associated with an entity in a particular context [29]. For every entity, there can be multiple partial identities. Digital identities are not necessary the same as real-world identities since the digital identities represented online may differ f rom t he c haracteristics r epresented i n t he physical world.

There are various definitions of digital identity. These definitions a re m ostly i nconsistent a nd s emantics b ased. A definition founded on mathematical properties would help in providing a uniform definition of digital identity and reduce confusion [30]. A *domain* can be defined as the namespace in which an entity is represented and uniquely identified. Assume that $D$ denotes the set of domains and $d \in D$ defines the domain of a single organization whereas $U_D$ stands for the set of users in that domain. In [30] Ferdous et al. define $A_d$ as the set of attributes and $AV_d$ as the set of their values within $d$. Then they describe the attributes for a user in a particular domain $d$ as

$$atEntToVal_d = A_d \times U_d \longrightarrow AV_d. \quad (1)$$

The above function returns the corresponding value of the attributes in domain $d$. The function is a partial function as not every entity has a value for every attribute.

An *identifier* is a unique value used to distinguish an entity in a given a domain [30]. The identifier $i \in A_d$ in domain $d$ is defined as an attribute that always exists and is unique within the context. These conditions can be defined as follows:

(a) $atEntToVal_d\ (i, u)$ is defined for all $u \in U_{d.}$

(b) $atEntToVal_d\ (i, u_1) \neq\ atEntToVal_d\ (i, u_2)$ for all unique $u_1,\ u_2 \in U_d$

The partial identity of a user $u$, for a domain $d$, can then be described as

$$parI\ de\ nt_u = \{(a, v) | a \in A_d, atEnttoVal_d(a, u) = v\}, \quad (2)$$

which can be rewritten as

$$parI\ de\ nt_d^u = \{(i_d, v_{i\ d}), (a_1, v_1), (a_2, v_2), (a_3, v_3), \ldots, (a_{n-1}, v_{n-1})\}, \quad (3)$$

where the $i_d$ and $v_{i\ d}$ represent the identifier and its corresponding value in domain $d$.

Finally, the complete identity of an entity is defined as the union of all their partial identities in all domains. This can be defined as

$$i\ de\ nt^u = \bigcup\{(d, parI\ de\ nt_d^u) | d \in D \text{ and } u \in U_d\}. \quad (4)$$

In [31] Ferdous et al. provide a more elaborate mathematical model for digital identity. By formalizing the con-cepts related to digital identity using mathematical properties, it is possible to describe and conceptualize the notions in a more formal and rigorous method and to capture the associated fundamental properties. In the fol-lowing section, we will describe the architecture of the identity and access management system.

*3.3. Architecture of Identity and Access Management.* An identity and access management system is a collection of tools, processes, and policies used to manage individual identities, their authentication, authorization, roles, and privileges, within an organization or across boundaries [19, 32]. An IAM system facilitates the administration of identities belonging to entities of an organization or a network. It keeps track of the roles and privileges associated with individual identities and provides the decision makers with means of control access to sensitive resources. Being an $8 billion industry in 2016 and reaching $20 billion by 2020 [33], the goal of an IAM system is to increase security and productivity and decrease downtime, cost, and repetitive tasks such as user creation and deletion. As the famous New Yorker cartoon "on the Internet, nobody knows you're a dog" states [34], there is a lack of adequate identity component on the Internet which had led to a huge opportunity lost, as having accurate and trustworthy identity of users is crucial to the growth of the digital economy.

To accurately define identity and access management, it is important to begin by describing a number of fundamental terminologies.

Subjects are entities, such as people or things that are under consideration [28]. A subject is able to access an *object*. An object is a passive entity that is being accessed by a subject. An example of an object is a particular online service such as a bank account.

As stated, *identifiers* are labels given to subjects. Identifiers are used to keep track of the information known about the subjects in a specific context. Examples of identifiers include social insurance numbers in Canada, social security numbers in the United States, and driver's license numbers. An entity may have multiple identifiers.

*Attributes* describe the characteristic of an entity. An attribute is a distinct and measurable name-value property belonging to an entity in a given context. The value of an attribute may be used to identify the entity, albeit the identification may not be unique to the entity [30].

Attributes can be issued by the entity itself or by another entity issuing them. Political affiliations, religious opinions, and gender are attributes that can typically be defined by the entity itself. There are also attributes assigned by one user to another. Attributes that can assist in the revealing or tracing of an individual's identity are known as Personal Identifiable Information (PII) [35].

A statement about an entity is known as a *claim*. Kim Cameron defines claim as "an assertion of the truth of something, typically one which is disputed or in doubt" [28]. A *credential* is a set of one or more claims made by the same entity. A *verifiable credential* is a set of tamper-evident claims and metadata that cryptographically prove who issued it [36].

Financial institutes, government agencies, and telecommunication companies are among potential trusted parties able to issue claims. The term *trust* is defined as "the firm belief in the truth or ability of someone or something to perform a task in a reliable matter, in a specific context" [37].

The main actors participating in IAM operations are shown in Figure 1. The *user* is a natural person with at least one digital identity that wishes to conduct a transaction [38]. An *identity provider* (IdP) is a special type of service provider that manages identity information. It is responsible for the creation, maintenance, and deletion of identity information belonging to the users. An IdP provides user authentication on behalf of the service providers [38]. A *relying party* (RP) or *service provider* (SP) is an entity that decides to provide its services, based on the information provided by other parties such as an identity provider and the user itself [38].

*3.4. Operations of Identity and Access Management.* Identity and access management systems provide a number of fundamental operations. As shown in Figure 2, they include identification, verification, authentication, and authorization.

Identification is the first step towards recognizing the identity of the user interacting with another entity such as a service provider. Identification occurs when a *subject* such as a user or a machine claims an identity.

This process is typically accompanied by a username, a unique ID, or anything that can uniquely identify the subject.

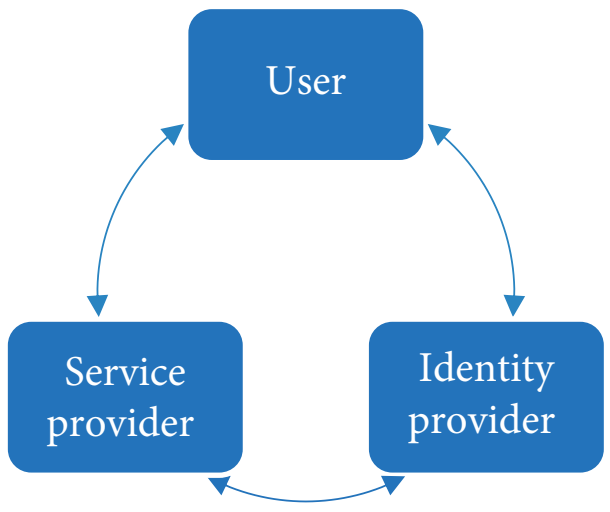

Figure 1: Identity management actors.

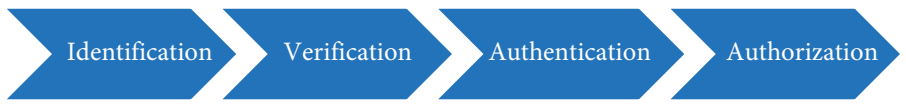

Figure 2: Identity management operations.

*Zooko's triangle* published in an article by Zooko Wilcox-O'Hearn in 2001 describes the three design decisions of an identification system to generate user identifiers. These design decisions are *secure*, *decentralized*, and *human meaningful*, out of which only two are likely achievable [39].

The *identity verification* operation, also known as identity proofing, is a relatively newer element of this list. Identity verification is the process by which specific identity attributes are truthfully associated with the entity they represent.

The *authentication* is verifying the identity of an entity such as user, process, or device, often as a prerequisite to allowing access to a system's resources [26]. This operation is typically performed by the entity presenting a token, password, or other forms of information such as biometrics prior to accessing the recipient's protected resource.

The *multifactor authentication* (MFA) is an authentication model that relies on more than one distinct authentication factor for a successful authentication. Common authentication factors involve three basic factors: first factor authentication corresponds to something the entity knows. A common instance of knowledge factor is a password, or a pin number known only by the entity and the recipient. The second factor authentication is attributed to something the entity possesses. Proving the possession of a mobile phone or a hard token is an example of this form of authentication. Third factor authentication represents what the entity is; this is typically achieved through biometrics such as fingerprint scan or facial recognition. Biometric authentication has been researched extensively in the recent years. There are various established and emerging standards in the context of biometrics authentication. The FIDO alliance [40] and The IEEE 2410 Biometric Open Protocol Standard (BOPS) [41] provide a standard for interoperability and open communication among the servers, browsers, and mobile devices to streamline biometric-based authentication. In 2018, the Rebooting the Web of Trust (RWoT) [42] proposed the six principles as guidance for using biometrics for self-sovereign identity models [43]. Lastly, the fourth factor authentication is commonly attributed to authentication based on contextual information such as time, location, behavioral patterns, and the entity's social reputation.

*Continuous authentication* is now a common industry concept. The purpose of continuous authentication is to provide unobtrusive and continuous user authentication, behind the scene, and without inconveniencing the users. The user may have zero to limited knowledge of the authentication process happening, while they interact with the service provider. Continuous authentication can be a supplement to the more traditional authentication forms to better guarantee the established identity of the user who has been authenticated at the beginning of their session. Moreover, the continuous authentication process provides a level of identity assurance that may not be present with traditional authentication systems that take place only during the initial user interaction. Continuous authentication actively monitors the behavior of the user to ensure the behavioral patterns match those of the expected and previously calculated authorized user. If there is any deviation, a red flag may be raised by the service, which may prompt the user to reauthenticate with their credentials.

*Authorization* is the process of granting access rights to an entity such a user or a program. The authorization process determines what the entity can do against a target resource.

A combination of identification, authentication, authorization, and audit processes working in tangent to ensure only authorized entities are allowed to access the resources is called *access control* [44]. There are many common access control models. They include discretionary access control (DAC), mandatory access control (MAC), identity-based access control (IBAC), label-based access control (LBAC), policy-based access control (PBAC), role-based access control (RBAC), capability-based access control (CBAC) [45], and finally, attribute-based access control (ABAC). Midhun et al. provide a summary of various access control models suitable for web applications [46].

*3.5. Evolution of Identity Management Models.* To address the evolving identity management requirements and provide minimal friction for users attempting to access online services, various identity management models have been designed [47, 48]. Figure 3 depicts the evolution of identity management models.

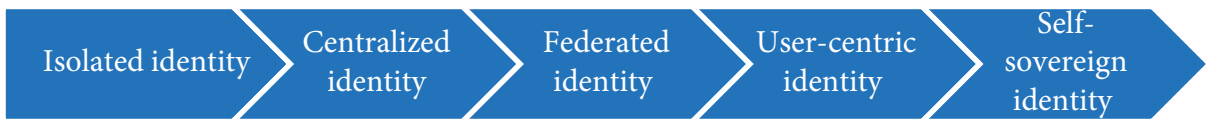

FIGURE 3: Evolution of identity management models.

The *isolated identity model* is one of the initial and more primitive IAM models in which the service provider is responsible for all identity management operations of its users. The major drawbacks of this approach are the large number of credentials that must be maintained by each user, and the risks associated with the storage of large amount of user data by every service provider [47].

Related to the above model, there is the *centralized identity model* in which the identity provider and the service providers are decoupled yet managed by the same organization. Every user interaction with the service providers must be authenticated through the central identity provider [47, 48]. The centralized model does not indicate a centralized IT infrastructure, but rather the centralized management of identity data, including the user's unique identifier, and all identity management operations such as issuance, authentication, and handling of digital identity life cycle. The centralized model is susceptible to various security attacks. For example, the disclosure of a user's credential is sufficient for giving access to all other services associated with that user. The centralized approach of this model is not an efficient and ideal approach for Internet users or users of significantly large organizations. Examples of this model are Kerberos [49] and public key infrastructure (PKI) [50] based architecture such as one-tier hierarchy and other hierarchy options that rely on a single root central authority (CA) to issue, maintain, and revoke digital certificates [51]. A CA can misuse certificates, issue invalid certificates [52], or become the victim of a security breach leading to a wide-ranging ramification including fraudulent issuance of certificates and man-in-the-middle (MITM) and impersonation attacks [53–55]. When a CA is identified as compromised, all digital certificates issued by that CA are considered invalid. As a result, a rapid software update must be rolled out to all clients relying on that CA as a root of trust to remove the affected root certificates. The PKI and decentralized PKI models will be explained in detail in Section 4.4.

In the *federated model*, a set of service providers and identity providers forms a trusted federation. This allows the user to have the option of authenticating through one of the federated identity providers to access any of the participating service providers.

The federated identity model allows a user to use a single set of credentials to authenticate with the identity provider, to seamlessly access any of the federated service providers. This is achieved by the identity provider providing an authentication token to the service provider. This feature is known as *single sign-on* [56].

Protocols such as Security Assertion Markup Language (SAML) [57] are leveraged to develop federated identity systems.

Often expressed as a little discussed yet significant iteration in IAM models, the *user-centric identity model* allows users to have some liberty on selecting their preferred identity provider and the identity attribute they like to share, along with the conditions under which those attributes should be shared [47]. In this model, the service providers and the identity providers may not always have a preexisting trust relationship. Examples of this model are the use of Facebook and Google login feature to access other online services. Protocols such as OAuth [58] and OpenID Connect [59] are the dominant technologies in this model. L'Amrani et al. [60] provide an in-depth evaluation of aforementioned identity models.

The most suitable model and configuration of the identity management systems adopted by any organization depends on various cultural, historical, jurisdictional, and technical determinants. Every type of identity management model is suitable for satisfying certain types of requirements [47].

*3.5.1. Self-Sovereign Identity.* The self-sovereign identity model is the next step in the continuum along the line of the centralized and federated identity models. While SSI has been gaining recognition in the recent years [61–63], it has its privacy roots in the 1970s by the introduction of Diffie–Hellman key-exchange protocol that allows users to protect their privacy in the digital environment using public key cryptography [64].

The term self-sovereign identity originated from a blog post in 2012 by Devon Loffreto to the Vendor Relationship Management (VRM) mailing list with the title "Sovereign Source Authority" [65].

Also known as *self-managed identity* and *user-controlled identity*, SSI is an identity management model in which the *identity holder* has broader control over their data and decides how and under what conditions their data should be shared with others. The SSI model is intended to preserve the right for the selective disclosure of the identity holder's data in different contexts. The identity holder is a role an entity might perform by possessing one or more claims. An identity holder is not always the subject of claims they are holding. In [66], Christopher Allen defines SSI as follows:

> "...The user must be central to the administration of identity. That requires not just the interoperability of a user's identity across multiple locations, with the user's consent, but also true user control of that digital identity, creating user autonomy. To accomplish this, a self-sovereign identity must be transportable; it cannot be locked down to one site or locale. A self-sovereign identity must also allow ordinary users to make claims, which could include personally identifying information or facts about personal capability or group membership. It can even contain information about the user that was asserted by other persons or groups."

Allen provides the following ten principles defining an SSI system [66]:

(i) *Existence.* "Users must have an independent existence."

(ii) *Control.* "Users must control their identities." They should have the liberty to manage their attributes in any way desired as they have complete authority over their identity data. Note that this does not stop entities from making claims about other entities. Also there should be a distinction between identifiers and identities. Identifiers must be managed by the user themselves. There should also be a distinction between control and ownership of the data. A user may possess control over an identity claim issued to them, but they may not be the owner of that claim. For example, a driver's license number is issued by the government and can be revoked by the government. However, the user should have the ability to control and share their driver's license number under their conditions.

(iii) *Access.* "Users must have access to their own data." There must be no gatekeepers or filters without the knowledge and approval of the identity owner. Note that this does not infer that any identity holder can modify all of the claims associated with their identity, but rather it means they should be aware of such actions. Moreover, this principle applies to every entity with respect to their own identity, and not to other's.

(iv) *Transparency.* "Systems and algorithms must be transparent." The algorithms, procedures, and processes involved in the model should be visible to the identity holders. Free, well-known, and open-source frameworks are a logical solution to this criterion, opening the algorithms to public examinations, evaluation, and avoiding vendor lock-ins.

(v) *Persistence.* "Identities must be long-lived." Identity data should be long-lived and persistent until they are explicitly decommissioned by the identity holder. Note that identity formation is an ongoing process and every identity management model must be designed to be adequately flexible to allow the identity holders to obtain, modify, or remove their identity data. For example, a person's name and address may change throughout their life. This does not mean that an identity holder controls all of the claims on their identity: other entities may make claims about the identity holder, but they should not be central to the identity themselves.

(vi) *Portability.* "Information and services about identity must be transportable." The identity holder should have the ability to seamlessly transport its identity data from one location to another. This is a necessary process for the longevity of the identity data. The identity data should not be locked into a single third-party entity if the third-party service provider has the best intentions of the identity holder.

(vii) *Interoperability.* "Identities should be as widely usable as possible." The identity data should be usable by as many entities as possible. The identity data should be consumable by a range of entities from different boundaries, jurisdictions, and architectures. This criterion supports the availability and durability of identity attributes.

(viii) *Consent.* "Users must agree to the use of their identity." The identity holder should have a clear understanding and provide their approval on how their identity data are used.

(ix) *Minimization.* "Disclosure of claims must be minimized." The least amount of identity data must be disclosed to accomplish the task at hand. This requirement is supported through the use of selective disclosure, range proofs, and other privacy-preserving techniques to ensure only the necessary data are disclosed.

(x) *Protection*. "The rights of users must be protected." The SSI model places identity holders in the center of its architecture. The right of identity holders must be preserved at all times. In cases where there is a conflict between the identity holder and the network, the network should err on the side of preserving the rights of the identity holder. Moreover, SSI architecture should be orchestrated in a decentralized form to avoid possible censorship and monopolies.

To avoid confusion, the distinction between the terms identity owners and identity holder in the context of SSI must be stated. An identity owner is an entity, such as a user, an organization, or a machine, that owns and issues a set of claims. On the other hand, an identity holder is an entity that holds in their possession one or more claims, and may not necessary be the owner of those claims.

The SSI model has been proposed as a step towards solving the challenges stated in Section 1 and the shortcoming of existing identity management models listed in Section 3.5. Section 8 provides an overview of academic work on SSI. The scholarly work around SSI is in early stages but growing. The majority of the engineering, model design, and collaborative advancements are taking place in the technical communities by means of discussions, projects, presentations, and whitepapers.

In the book "Self-Sovereign Identity: Decentralized digital identity and verifiable credential," authors provide a comprehensive definition of SSI, its building blocks, and governance framework [67].

In the book "Digital Identity Crisis," Rooly Eliezerov discusses the value of data and approaches to digital identity. He further elaborates on the key challenges of current digital identity models and the benefits of the SSI model [68]. Ferdous et al. [31] provide a survey of definition and analysis on SSI.

While growing rapidly, the SSI model has many challenges that need to be answered. Addressing these challenges accelerates its rate of adoption and use.

*3.5.2. Challenges of Self-Sovereign Identity.* The self-sovereign identity model is considered as the latest evolution of IAM models. It attempts to address a number of shortcomings found in the existing digital identity space. However, it also begets a unique set of challenges that warrant adequate research, exploration, and discussion. The following paragraphs discuss these challenges:

(i) *Standards for Data Management and Wallets.* Standard protocols, practices, and policies around user experience, data management, and data exchange should be carefully defined and implemented. While SSI supports an open ecosystem, there is value in consistent yet flexible user interactions, data management policies, and data presentation standards. Regardless of the direction taken to address this challenge, the approach should aim for a user-centric system aligned with Privacy by Design and Security by Design principles. The principles of Privacy by Design are described in Section 7.

(ii) *Key Management.* In traditional identity management models, the identity providers are primarily responsible for the management of identity data and secret keys and therefore must address the liabilities, risks, and technical requirements associated with that task. Conversely, in the SSI model, this responsibility and its associated risks are placed on the shoulders of the users. There have been numerous instances of users losing their cryptographic keys, resulting in the loss of valuable information and unrecoverable funds. Addressing the key management requirements in the SSI architecture is a fundamental step towards the mass adoption of SSI. Reliance on decentralized key custodians is one form of addressing the key management challenge [69].

(iii) *Consent.* As stated by Article 4 of General Data Protection Regulation (GDPR) [70], the consent given by the user must be meaningful, well-formed, unambiguous, specific, and freely given, specifying clear decisions. This process is not easy to implement and hard to verify in the current identity models. Moreover, requesting users to provide consent to many privacy policies and data sharing practices has led to what is known as *consent fatigue*, where the user is bombarded with privacy notifications. Ideas such as automated decision-making and response by digital wallets or agents representing the user, based on prior user decisions, should be entertained. Moreover, research around consent management, presentation, and enforcement is valuable.

(iv) *Access.* The backbone of many SSI systems is the distributed ledger technology (DLT). The concept of DLT and blockchain will be explained in the next sections. Certain DLT systems are public, allowing any entity to read or write to the ledger, while others are permissioned and allow only a selection of authorized entities to read or write new records into the ledger. If not carefully designed, the permissioned approach possesses the risk of forming centralized architecture similar to an oligopoly among the few authorized entities [71]. On the other hand, the permissionless and public model may be vulnerable to attacks common in various open DLT architectures [2].

(v) *Accountability and Governance.* It is important to articulate the policies and procedures around identifying and addressing malicious behavior and dishonest entities. It is also important to realize and articulate the correct degree of decentralization needed to support the vision and the requirements of SSI. Certain identity management operations such as identity claim issuance, identity lookup, and secure storage of data may rely on some degree of centralization and dependence on trusted intermediaries. Some implementations of SSI place significant power in the hands of a select few trusted entities that must comply with a common contractually binding trust framework, potentially making these entities the weakest point of the network. While other implementations of SSI aim for a more decentralized, programmable, and machine-readable governance framework. The latter form of governance too has shown to suffer from various flaws in the past [72]. Therefore, efforts to determine the correct balance between centralization and decentralization is an important topic.

(vi) *Trust in Data.* While there may be trust in the underlying SSI network as a secure, robust, and decentralized platform, the methods to form trust among the entities, and the trust in data including the verifiable credentials exchanged must be carefully designed. The authentication and data validation may need to be done through a trusted authority and outside of the blockchain network.

(vii) *New Technology Adoption.* As a new identity model, SSI requires various modifications to the existing system architectures. An important step towards the success of SSI is the discussions around the suitable technology stacks, deployment practices, and operational procedures. Particular attention must be given to the user experience, including the user interactions from the operator's perspective. Proper design steps must be taken to avoid the fate experienced by many other good innovations such as Pretty Good Privacy (PGP) [73], which while is a useful technology it has not met the expected broad use.

(viii) *Investment and Commercialization.* As a relatively new venture with a growing ecosystem but with limited knowledge on the revenue model, unknown user acceptance, and utilization and unknown risks, any entity intending to adopt SSI must design a strategic plan that supports the investment and risk involved in the deployment and operation of such system. The SSI economic model may lead to the chicken and egg problem where user adoption depends on the support of the service providers and vice versa.

An adequate response to the aforementioned challenges is the premise for a strong acceptance and support of the SSI model. Section 9 provides more perspective on existing SSI challenges.

In the next section, we delve deeper into the architecture of SSI and explain its building blocks.

## 4. Architecture of Self-Sovereign Identity Model

We begin this section by describing the key building blocks of SSI, including decentralized identifiers, verifiable credentials and decentralized public key infrastructure (DPKI), blockchain, and verifiable data registry, along with technologies such as identity hubs and agents which help realize the architectural requirements of SSI.

*4.1. Decentralized Identifiers.* Developed by World Wide Web Consortium (W3C) [74], the decentralized identifiers (DIDs) [75] are a key component of the SSI model. The DID specification defines a globally unique and cryptographic identifier scheme similar to the universally unique identifier (UUID) scheme [76] with a number of differences. First, the DID specification does not rely on a centralized authority to manage the identifiers (hence called decentralized identifiers), but rather they can be managed using a decentralized infrastructure such as DLT. Second, the DID addresses have cryptographic properties. The DID addresses are generated based on cryptographic key pairs. The ownership of a DID can be proven using cryptographic proofs such as digital signatures. As shown in Figure 4, a DID address consists of three parts. Every DID has the following format:

<Scheme>:<Method>:<Method Specific Identifier>.

The first part of each DID is the DID scheme, followed by the DID method. The third part consists of an identifier in the context of a DID method. Each DID address resolves to a *DID descriptor object* or DDO, also known as a *DID document*. A DDO is a machine-readable JSON-LD [77] document with various information about the DID subject. These include cryptographic public keys, service endpoints, authentication parameters, timestamps, and other metadata. The service endpoints provide a publicly available address for establishing connection with the DID subject. A *DID subject* is an entity identified by a DID and described by a DDO. A *DID record* consists of a key-value pair of a DID and a DDO. Figure 4 depicts a sample DDO document. An entity may have multiple key pairs and DIDs and use a different DID for each interaction.

The DID method defines the specific methods a DID scheme can be implemented on a particular DLT or network. This includes the CRUD operations of creating, reading, updating, and deleting of DID records. As of now, there are 90 registered DID methods [78]. These methods include Bitcoin, Ethereum, Sovrin, Interplanetary File System (IPFS) [79], and Veres Ones [80].

The DID specification distinguishes three types of DID [81]. The *Anywise DID* or *Public DID* is a DID intended for use with an unknown-able number of parties. The *Pairwise DID* is intended for interactions where the DID should only be known by its subject and exactly one other entity such as a service provider. Lastly, the *N-wise DID* is intended to be known by exactly $N$ number of entity including its subject.

A *universal resolver* [82] is a DID resolving system that supports multiple decentralized identifier systems. For a DID system to support the universal resolver, it must implement a DID adaptor to interface the universal resolver with the system-specific DID methods.

The *DID Auth* protocol allows an identity owner to use their client application, such as their mobile device or browser, to prove to a service provider that they are in control of a DID.

```
did:example:123456789abcdefghi
{"@context": "https://w3id.org/did/v1",
"id": "did:example:123456789abcdefghi",
"publicKey": [{
"id": "did:example:123456789abcdefghi#keys-1",
"type": "RsaVerificationKey2018",
"owner": "did:example:123456789abcdefghi",
"publicKeyPem": "Public key goes here.."
}],
"authentication": [{
// this key can be used to authenticate as DID ...9938
"type": "RsaSignatureAuthentication2018",
"publicKey": "did:example:123456789abcdefghi#keys-1"
}], "service": [{
"type": "ExampleService",
"serviceEndpoint": "https://example.com/endpoint/8377464"}
]
}
```

Figure 4: Sample decentralized identifier and DID document.

DID Auth protocol relies on a challenge-response cycle that is customized depending on the situation. This protocol can replace the use of username and passwords as a form of authentication and enables an authenticated communication channel between an identity owner and a service provider.

The *DID Comm* is a DID-based protocol by which two or more SSI entities can communicate privately and securely in a peer-to-peer manner. The protocol relies on DID and supports mutual authentication between the parties [83].

*4.2. Verifiable Credentials.* The *verifiable credentials* (VCs) [36] are a specification developed by the W3C Verifiable Credentials Working Group. A verifiable credential is an interoperable data structure suitable for representing cryptographically verifiable and tamper-proof claims. Figure 5 obtained from [36] displays the key roles within the verifiable credentials' ecosystem.

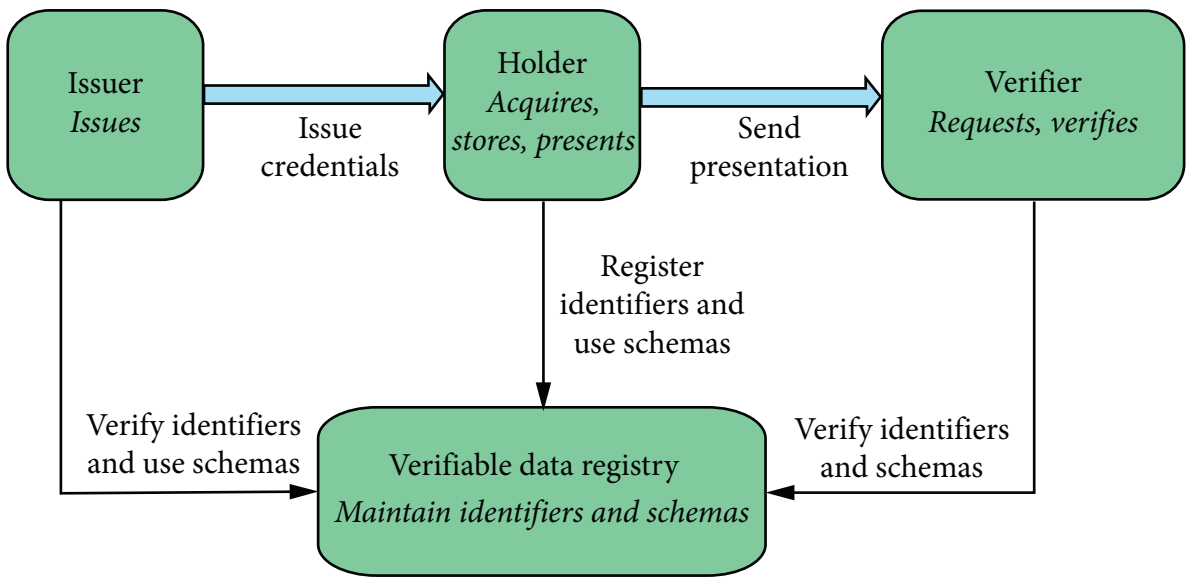

Figure 5: Verifiable credentials actors.

A *holder* is an entity that controls one or more verifiable credentials. An *issuer* is an entity that creates new verifiable credentials. A bank and a government agency are examples of credential issuers. A *verifier* is an entity that obtains one more verifiable credentials to verify. An e-commerce website expecting credentials from their customers is an example of a verifier. A *verifiable data registry* is responsible for mediating the creation and verification of identifiers, keys, verifiable credentials schemas, and other relevant data required to use verifiable credentials. A verifiable credential consists of various elements. They include the URI of the subject, the URI of the issuer of the claims, and the URIs that uniquely identify the credential. In addition, a VC includes claim expiration conditions and cryptographic signatures. A URI can be a DID. The W3C Verifiable Credentials Working Group has also defined the concept of *Verifiable Presentations* (VPs). Verifiable presentations define the methods by which VCs are signed and presented by the holder. A VC or a VP can be described with JSON-LD, JSON [84], or JSON Web Token [85].

*4.3. Decentralized Public Key Infrastructure.* The public key infrastructure consists of a set of services, tools, processes, and technologies which facilitate the performance of cryptographic operations based on public key cryptography. The most commonly used PKI certificate model is known as PKI X.509 or PKIX. In this model, the central certificate authorities (CAs) create the X.509 digital certificate. A certificate binds a public key to a particular identity [86]. Along with certificate authorities, the Domain Name Server (DNS) registrars and Internet Corporation for Assigned Names and Numbers (ICANN) [87] were created to facilitate the management and resolution of online identifiers and addresses. The governance model utilized by the above central authorities runs the risk of placing the control of identity data and decision-making in the hands of a small set of central authorities, leading to the possibility for those authorities to misbehave or become victims of security breaches [88]. Recent development in the formation of the Post-Transition IANA (PTI) entity as part of the transition stewardship program within ICANN, to perform the responsibilities of Assigned Numbers Authority's (IANA) [89, 90], and other advancement in introducing broader community-driven decision-making within ICANN have brought forth the hope of better accountability and democratic decision-making in this space [91].

A well-known implementation of a decentralized trust model, before the emergence of blockchain, is the Pretty Good Privacy protocol. Proposed by Phil Zimmermann and contrary to traditional PKI models, the PGP trust model relies on a web of trust. The popularity and growth of PGP were inhibited by various unanswered usability and key management challenges [88] in addition to the security concerns associated with the use of long-term keys.

Proposed by the Rebooting of the Web of Trust (RWoT) [42], the decentralized public key infrastructure (DPKI) provides a decentralized trust architecture in which no single entity can compromise the security and integrity of the system as a whole [51]. In contrast to traditional PKI, the DPKI does not depend on central certificate authorities or registration authorities (RAs). The dependence from these central entities can be eliminated by relying on a decentralized platform as the initial root of trust.

The DPKI architecture can be realized with the help of a decentralized key-value data store such as blockchain. By relying on the immutability of the blockchain to store the public keys, it is possible to guarantee that the data cannot be deleted or modified by anyone once it is written. The cryptographic keys can be found by anyone who has access to blockchain. No CA is not responsible for their storage or verification. This addresses some of the privacy and security concerns associated with a central point of authority. Because of the transparent and open nature of public blockchains, any changes to the data are auditable by all members of the network. Note that we must distinguish the trust placed on the blockchain system as a transparent, immutable, and secure storage platform, and the trust placed in the data itself, such as the public keys stored on the blockchain. While a verifier can rely on DPKI to allow a particular entity to authenticate itself and claim the possession of a particular identifier a nd p ublic k ey, t hat d oes n ot t ranslate t o the verifier trusting the entity's identity. The identification of an entity is achieved through the exchange of verifiable credentials. Although verifiable c redentials c ontain digitally signed identity claims, ultimately it is up to the verifier to trust and accept or reject those claims based on its own policies and trust model.

The S idetree p roject p roposed t hrough Decentralized Identity Foundation (DIF) [92] is an emerging protocol for creating scalable DPKI networks that can run on top of any existing decentralized anchoring system such as Bitcoin, while being as open and public as the underlying system they utilize [93]. Sivakumar and Singh [94] developed a blockchain-based DPKI by which key management operations are addressed through the use of smart contracts.

It is important to note that not all security breaches and concerns around traditional PKI are due to its fundamental architecture, but rather because of inadequate enforcement of security controls in a PKI deployment, or the lack of proper key management by users who engage with the system. Similar issues can be found in DPKI models which offer h igh d ata r esiliency b ut c an b e v ulnerable t o various attacks. Because of the role the CA plays, popular public CAs have adopted high levels of operational security and go to a great length to protect sensitive data. Decentralized PKI systems must also follow strict security measures. While some aspect of DPKI may rely on blockchain, many tasks such as cryptographic signing operations require secret keys which must be performed in a secure, central, and off-the-blockchain environment.

In addition, while DPKI systems offer a transparent and open architecture, the CAs in the traditional PKI model have also taken steps towards better transparency by offering services such as certificate t ransparency l ogs [95], through which all issued certificates are published on the CA website.

M any PKI enterprise systems have been deployed in various industries including banking, health, and defense. Due to the rapid growth of PKI ecosystem, deployment and management of PKI systems have become easier. There are now many on-premise and cloud-based PKI options available in the industry. For private ecosystems where certificate issuance and renovation detail of the parties must remain private, the cost of deploying a private DPKI with a large number of nodes may not be financially feasible.

*4.4. Blockchain and Distributed Ledger Technology.* The blockchain technology and its superset, the distributed ledger technology , provide a cryptographically secure, decentralized, and distributed database of information. The blockchain underpins the vast majority of cryptocurrencies, such as Bitcoin [2]. The blockchain disintermediates the central authorities, allowing the transactions to take place in a peer-to-peer method without the reliance on a trusted central party. Certain blockchain implementations such as Ethereum [29] support *smart contracts* capabilities. Smart contracts are distributed applications executed by the participating blockchain nodes.

The applications of blockchain have surpassed beyond the financial use cases and grown into other industries such as Internet of Things, smart homes, healthcare, supply chain management, energy, legal, voting, storage, corporate registry, and identity management. As the most well-known form of DLT, the blockchain technology became known after the introduction of the Bitcoin white paper in October of 2008 by a person, or persons of the name Satoshi Nakamoto. The Bitcoin protocol was officially launched on the third of January 2009. Bitcoin is by far the most well-known and longest running application of blockchain technology.

A blockchain system consists of an immutable ledger of cryptographically hashed transactions. A group of transactions collected together forms a block. Each block contains a header that references the cryptographic hash of the previous block in the blockchain. Figure 6 represents a chain of blocks, where each block references the block before it. Gao et al. [96] provide a comprehensive description of the blockchain system, its major components, algorithms, and challenges.

Blockchain systems can be divided into two categories of public and private, and two types of permissioned and permissionless. A public blockchain is completely available to all public members to read, while a private blockchain is accessible only to the authorized entities. Being a public network, PII information is generally not added to these systems. A permissioned blockchain provides read access to all parties; however, writing permission is only granted to a subset of entities. Conversely a permissionless blockchain allows for anyone to write to the blockchain. Bitcoin and Ethereum are examples of public permissionless blockchains as they are open to everyone to read or write. Hyperledger Fabric [97] is a popular private blockchain, while Hyperledger Indy [98] is an instance of a public permissioned DLT.

The consensus protocol is an integral component of blockchain. It ensures that only valid transactions are added to the blockchain. There are a large number of consensus algorithms [99] available. These algorithms date back to 1982 with the introduction of a paper by Leslie Lamport et al., describing the challenges and solutions of reaching Byzantine Fault Tolerance (BFT) [100]. While the consensus among known parties may be achieved through BFT and its variants, public blockchains such as Bitcoin where participants are not trusted must rely on other classes of consensus algorithms. Bitcoin relies on a consensus protocol known as *proof-of-work*. In this protocol, the participants of the network are required to solve a difficult mathematical puzzle in order to prove the validity of the transactions they intend to submit to the network. Bach et al. [101] and Du et al. [102]

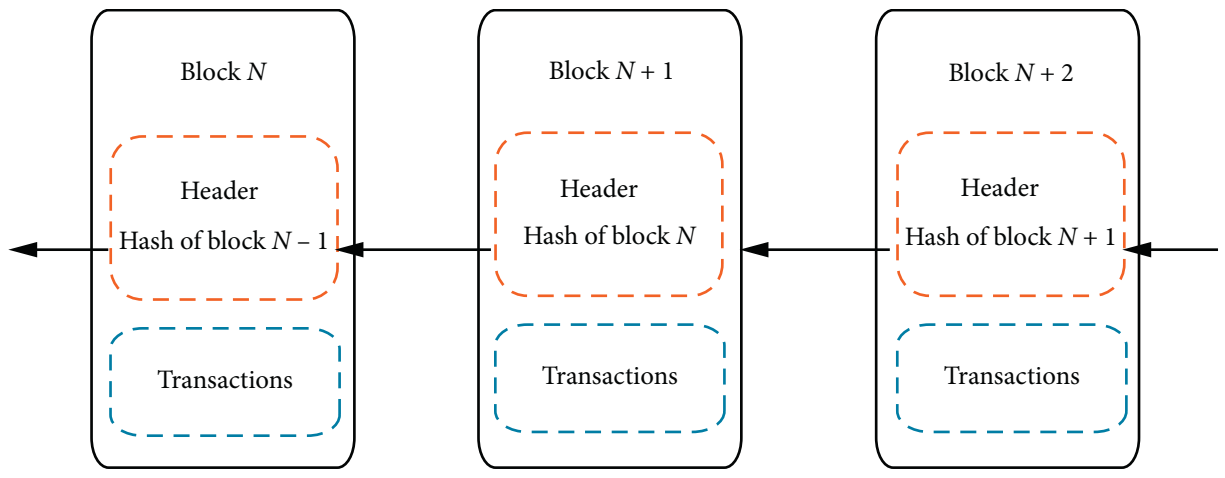

FIGURE 6: Blocks in the blockchain architecture.

provide a comprehensive overview and comparative analysis among various consensus algorithms.

Blockchain provides a framework of trust and the opportunity for data storage and processing to be diffused among numerous entities. By leveraging this architecture, the need for the continuous availability of a central identity providers may no longer be necessary.

The advantages of using blockchain to achieve IAM requirements are four folds. First, the decentralized nature of blockchain allows data and decentralized identifiers to be stored in a ledger and not with a single central entity. Second, as blockchain is an immutable, the transactions can only be appended to the ledger and its integrity can be verified by all members of the network.

Third, in the case of a public blockchain, all records are always visible to all.

Finally, due to the distributed nature of blockchain, it is very difficult to influence the availability and integrity of the data.

Various types of operations such as registration of decentralized identifiers, notarization of identity data, and storage of access and consent records can be shared on blockchain. The decentralized nature of the blockchain allows for entity identifiers and cryptographic keys to be registered on an immutable, decentralized system accessible by any entities that need to obtain the authentic public key of an identifier. This is in contrast to the public key infrastructure, where the public keys are exchanged directly through a presumably secure channel, with strong dependence on certificate authorities to ensure the integrity and authenticity of the public keys. In more recent SSI architectures, different information, such as entity identifiers, is no longer added to the blockchain system, as they are considered PII information. Furthermore, by leveraging smart contracts on blockchain, it is possible to develop sophisticated and decentralized identity management platforms on the blockchain.

As discussed in Section 4.3, while a blockchain system provides integrity, immutability, and authenticity of cryptographic keys, it fails to prove the trustworthiness of cryptographic keys without the existence of a proper identity layer, governance framework, and trust policy in place. Moreover, in the context of public blockchains, privacy concerns around identity data and issues around data correlation are important topics of discussion.

*4.5. Verifiable Data Registry.* The goals of self-sovereign identity can be realized with the help of a trusted database platform such as a blockchain playing the role of a *verifiable data registry*. A verifiable data registry is responsible for mediating the creation and verification of identifiers, keys, verifiable credentials schemas, revocation registries, issuers' public keys, and other relevant data required to generate or use verifiable credentials. Note that other forms of databases, such as government ID database can also be considered as a verifiable data registry. There can also be more than one type of registry utilized in an SSI deployment.

*4.6. Agents.* An agent is a piece of software component assigned to act on behalf of an entity [103]. Agents serve a multitude of purposes. In the context of an SSI network, an agent is the main gateway between an entity and the rest of the network by providing a network address for other entities to connect through. Generally, an agent component has access to a digital wallet in order to store cryptographic keys and to perform various cryptographic operations. The agent component ensures resource availability, increased privacy, and auditing capability. In addition, it provides effective data management services and supports automated actions on behalf of the user. The drawbacks of agents include the trust placed in them to handle sensitive data on behalf of the entity [71].

There are two broad categories of agents, namely, *edge agents* and *cloud agents* [104]. An edge agent is an agent service running in the immediate vicinity of the user. This includes the user's mobile device or browser. The edge agent interacts directly with the user and has the ability to provide more direct and immediate data control to its user. An edge agent is required to be available at all times in order to respond to any service requests. A cloud agent, on the other hand, is an agent service hosted on the cloud. Generally, a cloud agent has more restrictive access policies than an edge agent. Generally, a cloud agent has a fixed and public address, making it susceptible against identification through correlation attacks. The cloud agent is capable of relaying all requests to its associated edge agent running on the user's device. Service providers that provide cloud-based agents as a service are called *agencies*.

Various efforts in standardizing the agent's interface and communication protocol are underway [105]. Smart agents are advanced agents with the ability to respond to various requests with minimal user involvement. Smart agents can be equipped with artificial intelligence to improve their decision-making capabilities. Google Assistant and other smart devices can enhance the user experience of agents [106].

*4.7. Identity Hubs.* Proposed by Decentralized Identity Foundation (DIF) [107], an *identity hub* is a secure data storage solution that provides a safe, highly available, and user-centric depository of user data. Identity hubs support interoperable and standard protocols for effective management and exchange of identity data and cryptographic information. Identity hubs support synchronization protocols to coordinate user data among multiple hubs and are required to support data exportation capabilities to allow the user's data to be exported in an interoperable format at the user's request. Authentication with an identity hub is supported through various protocols such as DID authentication scheme. An alternative yet similar solution to secure storage of identity data is known as the Solid project [108]. Solid provides a decentralized data store called a Pod, similar to a container, for storing identity data of the users. They support file- and graph-based data structures and can be hosted by a Pod provider or by the user themselves. As of now, there are four Pod providers. The Solid project supports OpenID as its main authentication protocol.

*4.8. Digital Wallets.* A digital wallet is a software or hardware component responsible for the storage of identity data and cryptographic keys in a secure and privacy-respecting method. In addition, often in concert with an agent, a wallet can perform various cryptographic operations such as key generation, signature creation, signature verification, and key backup and recovery. In order to comply with existing privacy and auditing requirements, a wallet may support the tracking of consent requests and responses, as well as all user interactions. A digital wallet can be deployed in many forms. It can be a program on the user's mobile device or browser or provided as a software as a service via the cloud. Lastly, a user may have multiple wallets.

## 5. Privacy Engineering

Privacy engineering covers the tools, processes, techniques, metrics, and taxonomy to implement Privacy by Design. It ensures that privacy considerations are integrated into the system design. NIST defines privacy engineering as "A specialty discipline of systems engineering focused on achieving freedom from conditions that can create problems for individuals with unacceptable consequences that arise from the system as it processes PII" [109].

Privacy engineering can supplement and extend SSI by addressing various critical requirements such as consent and data minimization. Because privacy engineering is vast and rapidly evolving field, this section only describes a selection of highly discussed topics including Trusted Execution Environments, Secure Multiparty Computation protocols and Ring Signatures, Threshold Secret Sharing, and Zero Knowledge Proofs protocols.

*5.1. Trusted Execution Environments.* Smartphone ownership is accelerating globally along with the growth of their capabilities and the ability to run more applications. This contributes to the increasing number of smartphone vulnerabilities. Trusted Execution Environment (TEE) can provide a method of protecting mobile applications and user's data from unauthorized access. Digital wallets and agents play a vital role in SSI architecture. The wallet must securely store user's cryptographic keys and identity attributes. Wallets can leverage TEE to maintain data confidentiality and secure code execution. Trusted Execution Environment is a tamper-resistant, secure, and protected processing enclave inside a processor that runs in parallel, but isolated, to the main operating system running on the device [110]. A TEE component provides an enhanced level of code and data confidentiality and authenticity of executed code, and the integrity of the run-time states. Applications that run in a TEE zone have access to the full features of the device; however, they are isolated at the hardware level from the user-installed applications running on the main operating system. Intel's SGX, AMD SEV, and ARM's TrustZone are popular examples of TEE.

*5.2. Secure Multiparty Computation and Ring Signatures.* Ring Signature is a privacy-respecting digital signature protocol that provides the ability to prove the existence of a signer that possesses the signing key among a group of signers without revealing the signer's public key. Ring signatures make it possible to produce a valid signature without revealing the identity of the signer [111]. Ring Signatures and Secure Multiparty Computation (SMC) [112] supplement the SSI architecture by enabling joint computations while maintaining data confidentiality. They also allow for retaining more control over who receives the outcome of the computation.

Secure Multiparty Computation provides a privacy-preserving method for a group of entities to jointly compute a function over their input without the need to share their input with other parties, or with a trusted third party. A trusted third party is an entity trusted by other entities within a particular context. As an example, assume that there are three parties Alice, Charlie, and Bob with secret inputs $x$, $y$, and $z$ representing their respective salaries. To find out the highest of the three salaries without revealing to each other their salary, they engage in an MPC protocol. The output of the protocol is simply the highest value of the salaries which can then be used by each party to compare against. Mathematically, the problem of finding the maximum value translates to $F(x, y, z) = \max(x, y, z)$.

History of secure multiparty computation begins in 1986 when Yao proposed the two-party computation protocol using cryptography [112]. Goldreich et al. [113] extended Yao's idea by providing a polynomial-time algorithm to solve the multiparty (mental game) problem under the assumption of the majority of parties being honest.

Various applications of SMC such as the private information retrieval problem (PIR) [114] are introduced by Chor and Gilboa. The PIR describes a server-client architecture in which a client retrieves the $i^{\text{th}}$ bit of a binary sequence from the server without the knowledge of the server on which bit has been obtained. The server, on the other hand, does not reveal the entire bit sequence to the client.

Privacy-preserving data mining is another common application of SMC. The author in [115] introduces an approach in which two parties, each with a separate database, are able to jointly perform data mining operations on the union of their data sets without disclosing their database to one another or any third party.

Nair et al. [116] propose a secure information storage and retrieval model based on SMC. This model splits sensitive data among multiple databases and reconstructs the data by a computation agent on the client side.

Clifton et al. [117] describe a toolkit for various privacy-preserving SMC-based data mining operations including secure sum computation using a secret random number. However, in this protocol, two adjacent parties can collude to compute the random number. To avoid the risk of two parties colluding to determine the secret random number, Mishra et al. [118] propose a privacy-preserving k-secure sum protocol to allow the parties to compute the sum while keeping their individual data secret. This is achieved through the segmentation of each individual party and generation of random numbers within each segment.

Pathak et al. [119] extend on this work to propose a secure and privacy-preserving data mining protocol. In their proposal, each party breaks its data block into $k$ segments and distributes $k-1$ segments to other parties. Through a series of interchanges between the parties, the total sum of the data is calculated while ensuring the privacy of individual data.

Pinkas [120] provides a detailed definition of SMC and various other privacy-preserving data mining techniques. Finally, the Enigma project is a peer-to-peer network that relies on SMC and blockchain to provide secure computation while keeping data private [121].

*5.3. Threshold Secret Sharing.* Threshold Secret Sharing and secret segmentation refer to the distribution of secrets into multiple shares. These shares are then disclosed with multiple participants. To recover the secret, a predetermined number of shares must be retrieved from the participants. More formally, given a finite set $P$ of participants and a set $\Gamma$ of subsets of $P$, the secret threshold scheme defines a threshold $t$ where the secret access model is $\Gamma = \{A \subseteq P\colon |A| \geq t\}$. In these schemes, less than $t$ participants cannot recover the original secret. This scheme is also known as the $(t, n)$-threshold scheme.

The SSI architecture can take advantage of Threshold Secret Sharing protocol to offer secure and privacy-respecting backup and recovery of user's sensitive data, including secret keys [122].

In 1979, Shamir [123] and Blakely [124] proposed two variations of threshold-based secret sharing schemes. The scheme proposed by Shamir relies on standard Lagrange polynomial interpolation, while Blakley's proposal, also known as vector scheme, relies on the geometric concept of intersecting hyperplanes. An evaluation of Shamir's Threshold Secret Sharing scheme against Blakley's approach is discussed in [125].

Shamir's secret sharing algorithm is described as follows: user $V$ splits a secret $S$ into $P$ shares, such that any combination of less than $t$ shares cannot learn the secret $S$, while any combination of $\geq t$ shares can learn the secret $S$. To this end, the user $V$ constructs a polynomial $f$ of degree $t-1$ such that $f(0) = S$. For example, if at least 4 shares are required to reconstruct $S$, then the polynomial will be $f(x) = a_0 + a_1x + a_2x^2 + a_3x^3$. Next, user $V$ computes $f(x_i)$ for all shares $i \subseteq P$ and the resulting pairs $\{x_i, f(x_i)\}$ are distributed to each participant $i$. In order to reconstruct the secret $S$, the polynomial $f(x)$ should be recreated from the set of pairs provided by the participants using Lagrange polynomial interpolation. The major steps of Lagrange interpolation are as follows: given a polynomial $f(x)$ over of degree at most $l$ and given $C$ such that $|C| = l + 1$, then

$$f(x) = \sum_{i \in C} f(i)\delta_i(X), \tag{5}$$

where $\delta_i(X)$ is a degree $l$ polynomial such that

$$\delta_{i(j)} = \begin{cases} 0, & \text{if } i \neq j, \\ 1, & \text{if } i = j, \end{cases} \tag{6}$$

which can also be represented as

$$\delta_i(j) = \prod_{j \in C,\, j \neq i} \frac{X - j}{i - j}. \tag{7}$$

*5.4. Zero Knowledge Proofs.* Zero Knowledge Proof (ZKP) is a cryptographic protocol by which a *prover* is able to, with high certainty, prove the validity of a statement to a verifier without sharing the underlying information [126]. The prover is able to maintain its privacy while providing to the verifier enough information to be able to validate the truthfulness of a statement. The ZKP protocol enhances user privacy while maintaining the required institutional trust for the correctness of digital interactions.

ZKP protocols are used in various applications including privacy-respecting identity proving and authentication, as well as cryptocurrency transaction validation and private data mining.

ZKP augments SSI's data minimization requirements and empowers users by impeding the unnecessary collection of their data by a verifier. In the context of verifiable credentials, ZKP introduces the following key capabilities [36]:

(i) The ability to combine multiple VC from multiple issuers into a single verifiable presentation while avoiding sharing unnecessary VC or identifiers with the verifier, making it more challenging for the verifier to collude with any of the issuers.

(ii) The ability to selectively disclose the necessary claims with a verifiable credential to a verifier without requiring the need to obtain multiple atomic VC from the issuers. This allows the holder to only share the information needed by the verifier.

(iii) The possibility to produce derived verifiable credentials formatted according to the verifier's data schema, without the involvement of the issuer.

As an example, a prover is able to prove to a verifier that he is above the age of 19 without revealing his actual date of birth, or a prover can prove that she lives in a country within the European Union without revealing the name of the country.

Zero Knowledge Proofs were first published by Goldwasser et al. in 1985 [126]. After 27 years from the time the paper was published, Goldwasser et al. won the Turing award for their effort in ZKP.

A simplified form of ZKP is the public key cryptography, where the possession of a private key is proven by a challenge-response protocol. Notable works are Succinct ZK by Kilian in 1992 [127] which provides a simplified ZKP, and Pinocchio by Parno et al. in 2013 which applies ZKP to real-life use cases [128], as well as Groths's Groth16 proposed in 2016 [129], which significantly reduces the protocol's computational complexity. The Bulletproofs protocol by Bünz et al. [130] has contributed significantly to the progress of ZKP. Other notable ZKP protocols include CL-signatures [131], Fiat Shamir [132], zkSNARKs [133], and zkSTARKs [134]. IBM Identity Mixer [135] and Microsoft U-Prove [136] are also among discussed ZKP implementations.

ZKP protocol is a type of proof systems. Proof systems can be divided into interactive and noninteractive proof systems (NIPS). An interactive ZKP requires the prover and verifier to be present during the verification process through which multiple messages are exchanged between the parties. At the end of this process, the verifier will either accept or reject the proof provided by the prover. The exchange between the parties requires an active connection. Conversely, the noninteractive proof systems, first introduced by Blum et al. [137], do not require the concurrent availability of both parties. Huixin Wu et al. provide a survey of noninteractive ZKP systems and their applications [138]. Every ZKP protocol must satisfy these three principal properties:

(i) *Completeness.* If the input to a ZKP system is true, the proof must also be true. In other words, referring to the example above, if the date of birth attribute meets the condition of greater than 19 years, then the result of the system received by the verifier must be accepted (with high probability) as true.

(ii) *Soundness.* If the input to the ZKP system is false, then the proof must also be false. In other words, it is not possible for the prover to trick the system to convince the verifier when the input is false.

(iii) *Privacy.* The confidential input of the ZKP system must not be revealed to parties other than the prover.

There are a number of canonical examples illustrating ZKP protocol, including the Ali Baba's cave [139] and the two pens scenario [140].

A ZKP ecosystem consists of various roles and terminologies, which are described as follows:

(i) Attributes are units of information about the identity holder. Date of birth or nationality are examples of attributes.

(ii) Holder is the entity who possesses the attributes. These attributes are attested by one or more authoritative sources, also known as the issuer. The holder uses the attributes to generate proofs to prove some claims about the attributes.

(iii) Issuer is the authoritative party that attests to the validity of the holder's attributes. For example, a government agency may issue a certified date of birth attribute to the holder.

(iv) Verifier is the party that verifies the claims, by verifying the proofs provided by the holder.

A successful ZKP protocol must consider a number of criteria, such as scalability, interactivity, security and threat modeling, transparency, and quantum security. In order for a ZKP protocol to work effectively, it must provide sufficient privacy measures to all parties. For instance, holder and issuer should be able to maintain their anonymity while the confidentiality and untraceability of claims are still satisfied.

There is a growing interest in ZKP, and its use cases are growing. For instance, use cases around the proof of a customer's age at an establishment that serves alcoholic beverages, privacy-preserving voting and membership systems, and use cases around satisfying know your customer (KYC) and anti-money laundering (AML) requirements while maintaining client's privacy are actively discussed within the ZKP communities.

Many organizations [141, 142] are actively working on extending ZKP protocols. Many technical groups such as ZKProof [143] aim to advance the use of ZKP by creating formal frameworks and standards around the scheme and by promoting stronger community collaboration among different stakeholders including the academia.

Data minimization practices limit the amount of data shared to the minimum necessary to successfully accomplish a goal. Other approaches to preserving privacy recommended by W3C include selective disclosure and progressive trust. Selective disclosure is the ability for a user to have granular control over what information they share. Progressive trust is the ability for a user to gradually increase the amount of necessary data revealed as the trust is build or value generated [144].

While not covered in detail in this survey, other notable cryptographic protocols include homomorphic cryptography [145], differential privacy [146], identity-based encryption [147], and commitment schemes [148] which may supplement a digital identity or data management systems. Bernabe et al. [149] provide a comprehensive survey of privacy-preserving research solutions within the context of blockchain and self-sovereign identity systems. In the following section, we describe the noteworthy platforms, technical groups, and organizations that are driving SSI protocols forward.

## 6. Platforms, Technical Groups, and Technology Leaders in SSI

There is a rapidly growing number of SSI platforms, industry players, and technical groups committed to developing and improving the SSI and decentralized identity framework.

*6.1. Platforms.* Stacks, formerly known as Blockstack [150], is an open-source decentralized naming and storage platform built on the blockchain technology. Similar to Ethereum, Stacks supports decentralized applications; however, the computations of these applications are performed off-chain. Civic [151] is an Ethereum-based decentralized identity management platform supporting a built-in incentive program. Similar to many other platforms, Civic's users use a mobile application to manage their identity data. The mobile applications are not yet open-source. uPort [152] is an SSI framework based on the Ethereum public blockchain and Interplanetary File System (IPFS) [79].

Veres One [80] is a nonprofit identity project with the goal of addressing a range of existing identity challenges. Veres One supports a public permissionless network that relies on a blockchain system optimized for identity use cases. It uses verifiable credentials, decentralized key management, and DIDs to facilitate a decentralized identity management model. Similar to a number of other frameworks, Veres One relies on a board of governors, a selection of qualified members responsible for the governance of the network. Important entities include nodes that are in charge of computation and storage of the network, maintainers who are responsible for maintaining the Veres One software, and entities such as people and organizations which interact with the network to create and manage their DIDs. There is also a special role of the accelerator, responsible for increasing the speed of creation and management of DIDs.

The Hyperledger Consortium [153] is an umbrella project managed by the Linux Foundation. The objective of the Hyperledger project is to create, foster, maintain, and support open-source distributed ledger framework-related projects. Hyperledger houses many DLT projects including Hyperledger Fabric [97], a general purpose DLT framework. Hyperledger also manages various DLT tools to support the development of decentralized solutions. These tools include Hyperledger Explorer and Hyperledger Composer. Currently known as one of the most popular permissioned blockchains, Hyperledger Fabric is a general purpose blockchain implementation originally developed by IBM. Fabric is developed with Golang programming language and supports pluggable consensus component, and smart contracts (known as Chaincode), and privacy and confidentiality features required for enterprise and commercial use cases. Other notable projects within Hyperledger include Aries and Ursa. Hyperledger Aries is a blockchain-agnostic, off-ledger key and secret management system. It provides implementations of agents and ZKP-capable verifiable credentials and decentralized key management system. Hyperledger Ursa is a cryptographic library shared among many Hyperledger projects with the purpose of offering a secure, consistent, and easy-to-use cryptographic protocols and services.

Originally developed by Evernym [141], Hyperledger Indy [98] is a public permissioned DLT solution purpose-built for decentralized identity use cases. Built from the ground up using Python and Rust programming languages, with the goal of addressing SSI use cases, Indy supports decentralized identifiers, verifiable credentials, and Zero Knowledge Proofs, while adhering to Privacy by Design principles. As opposed to storing blocks, or smart contracts, as is the case with many DLT frameworks, Indy supports the storage of DIDs and DDOs and various public identity-related records such as public claims, revocation material, and consent records. No personal information is recorded on the ledger without the approval of the identity owner.

The Sovrin Network [103] is an open-source user-centric SSI framework. Sovrin is originally developed by Evernym and is managed by the Sovrin Foundation, a global nonprofit organization [154]. Based on the Hyperledger Indy, the Sovrin tech stack supports verifiable credentials, DIDs, agents, and Zero Knowledge Proof protocols. The Sovrin distributed ledger is governed by trust anchors. Nodes in the Sovrin framework are run by stewards. There are two types of nodes, namely, validator nodes and read-only observer nodes. Sovrin stewards include companies from many continents and well-known market players such as IBM and Cisco. Sovrin supports a native protocol token to produce incentives and simplify various types of identity transactions among issuers, verifiers, and holders. The Sovrin Trust Framework [154] addresses the governance and policy aspect of the Sovrin network. It describes the processes and policies required to participate in the network, in addition to introducing various roles including trustee, trust anchor, and stewards.

Jolocom [155] is another open-source SSI platform that uses Ethereum by default. It relies on DIDs and hierarchical deterministic (HD) keys to generate multiple identities from a seed master identity. Every DID generated on Jolocom resolves to a DDO stored on IPFS. The mapping of every DID to DDO is performed on the Ethereum blockchain. While Jolocom SDK interfaces with Ethereum (through ERC-725 standard), the protocol is open for other blockchain implementations. Similar to other platforms, Jolocom users utilize a mobile app to interact, create, manage, and share their data.

SelfKey [156] is another SSI network where personal data are closely controlled by the users. By relying on Zero Knowledge Proofs, the platform ensures only the minimum necessary amount of data is shared.

LifeID [157] is built on an open and permissionless blockchain technology. It utilizes DID to generate identifiers and allow its users to control their identity data. Furthermore, LifeID relies on ZKP to ensure data minimization. A user's identity data can be backed up on cold storage, or stored with other users, or a trusted organization. LifeID has a token named ID Token to support storage and exchange of value on the network.

Finally, Mattr [158] is another SSI platform based on verifiable credentials and DIDs. The Mattr platform supports extensible data formats and secure authentication protocols. Kuperberg [159] provides a survey of various emerging SSI implementations. Dunphy et al. [160] and Abraham [161] provide further detailed comparison and analysis of the aforementioned platforms.

*6.2. Technical Groups and Technology Leaders.* The development *f* SSI is made possible through the work of various technical groups that help design the requirements, specification, standards, and processes for SSI.

The World Wide Web Consortium [162] is an international community that creates open standards for the web. The W3C Verifiable Credential Working Group [74] is actively working on the verifiable credential data model and use cases while the W3C Decentralized Identifier Working Group [163] is focused on developing the decentralized identifier specification, including its method registries and use cases. The goal of W3C Credentials Community Group [164] is to explore and draft specifications around the creation, storage, presentation, and control of verifiable credentials for further standardization and prototyping.

The Rebooting of the Web of Trust (RWoT) facilitates discussions on identity-related topics with a particular focus on a decentralized trust-based identity systems. The RWoT organizes annual events with the goal of producing five technical white papers on identity- and trust-related topics that have the biggest impact on the future of identity. Various initiatives such as DIDs, DPKI, and JSON-LD are among the initiatives examined in RWoT.

The Hyperledger Identity Group [165] promotes discussions, research, and collaboration on management of digital identity data on decentralized platforms and in particular solutions related to Hyperledger.

The Internet Identity Work Group (IIW) [166] is another technical group that meets twice a year to discuss new ideas, solutions, and initiatives with the primary focus on identity.

The Decentralized Identity Foundation (DIF) [92] is a technical organization focused on developing foundational elements required for an open ecosystem for decentralized identity. The DIF provides technical specifications and reference implementations and assists in coordinating the industry leaders. Identity hubs and universal resolver are among the many projects incubated by DIF.

The Digital ID and Authentication Council (DIACC) [167] in Canada is a union of public and private sector players working together to create a trusted digital identity experience for Canadians. The Pan-Canadian Trust Framework (PCTF) spearheaded by DIACC is a framework consisting of the legal, business, and technical rules applicable to identification, authentication, and authorization, agreed on between participating public and private sector organizations across Canada.

Kantara [168] is a global nonprofit organization developing specifications such as *User Managed Access* [169] and *Concept Receipt Specification*. The ID 2020 [170] is a nonprofit public-private consortium committed to improving lives by providing digital identity to all undocumented people worldwide. The goal of ID2020 is to provide identification for all citizens of the world by 2030. Finally, initiatives such as One World Identity (OWI) [171] and SSIMeetUP [172] provide industry practitioners the latest ideas and trends about SSI.

Finally, the purpose of Trust over IP Foundation (ToIP) [173] is to provide an open and collaborative platform to define a complete architecture for Internet scale digital trust that combines both cryptographic trust at the machine layer and human trust at the business, legal, and social layers. The mission of ToIP is to combine available standards, protocols, and capabilities produced by standard development organizations and industry foundations such as W3C and DIF, to fulfill the requirements of an Internet scaled digital trust architecture. The ToIP is composed of four different working groups, namely, Governance Stack Working Group, Technology Stack Working Group, Utility Foundry Working Group, and Ecosystem Foundry Working Group.

In the following section, we provide a summary of current and emerging privacy laws and regulations that have strongly shaped SSI and other privacy-respecting identity models.

## 7. Regulations

In recent years, we see the growing interesting in developing more comprehensive privacy frameworks that place the users as first-class citizens.

Data privacy regulations in various shapes and forms have been around since the $16^{th}$ century [174] and continue to evolve to this day to meet the changing technological landscape and demands of the stakeholders. In this section, a number of relevant regulations and standards are presented. They include the General Data Protection Regulation (GDPR) [70], Payment Service Directive (PSD2) [175], Privacy by Design (PbD) [176], and Electronic Identification, Authentication, and Trust Services (eIDAS) [177].

The General Data Protection Regulation (GDPR) was introduced by the European Union and came into effect in May 2018 [70]. This regulation replaces the previous data protection directive from 1995. The goal of the GDPR is to harmonize the privacy laws across Europe and to protect consumers by providing them the necessary protection in handling their personal information. GDPR consists of 11 chapters, 99 articles, and 173 recitals, and it is by no means a small regulation. The SSI model is aligned with the objectives of GDPR as it gives individuals control over their identity attributes and allows a streamlined exchange of credentials among the identity holders and other entities of the network while maintaining privacy and trust.

The Payment Service Directive, effective from January of 2018, provides a harmonized set of rules across the EU that enables bank customers, both businesses and consumers, to allow third-party providers to manage their financial needs [175]. Online services such as personal accounting platforms, social media services, and e-commerce websites can take advantage of the data provided by the bank accounts to provide financial services to their users. PSD2 will push organizations to rethink how they manage the digital identity of their clients. The SSI model is a strong contestant in addressing the identity and consent management requirements brought about by PSD2, but further research is required.

Introduced by the European Union, the eIDAS aims at standardizing the electronic signature, electronic timestamps, and documents, as well as electronic identification of natural and legal persons [177]. The eIDAS aims at creating common ground among the various electronic identification systems developed individually by each EU state. Exploring the implementation of the SSI model while complying with eIDAS to facilitate the creation and management of interoperable identity schemes is a potential research opportunity.

Privacy by Design (PbD) is an engineering framework originally developed in the 90s by Ann Cavoukian, in which the importance of privacy is taken into account throughout the engineering process. The PbD has the following seven foundational principles, namely:

(a) Proactive not reactive, preventive, not remedial
(b) Privacy as the default setting
(c) Privacy embedded into the design
(d) Fully functionality and preserving positive-sum, not zero-sum
(e) End-to-end security and full life cycle protection
(f) Visibility and transparency
(g) Respect for user privacy and keeping it user-centric

The principles of PbD should be considered and manifested during the development of a digital identity solution, to ensure the privacy of the identity owners, while preserving the security and integrity of the system as a whole.

Observing the privacy landscape in the United States, it is possible to see the growing interest by various states in composing legislation to protect the privacy of their citizens. Notably, California has proposed the California Customer Privacy Act (CCPA) [178] which went into effect in January of 2020.

In the following section, we will survey the research landscape and summarize the relevant academic body of work.

## 8. Related Work

This section presents the related literature on SSI. The section is divided into two parts. The first part provides a breadth of existing surveys on SSI, and the second part provides a summary of original SSI-based proposals.

*8.1. Surveys.* Grüner et al. [179] assess the use of blockchain technology for IAM through the use of a decision model on a selective group of IAM implementations, namely, uPort, Sovrin, and ShoCard. Based on their research, both uPort and Sovrin are IAM models fit for blockchain, although because of Sovrin's strict trust framework, it can exhibit a more centralized model than uPort. The focus of this paper is more on defining and applying a decision model rather than providing an overview of SSI and its building blocks, or discussing privacy techniques.

Jacobovitz [180] provides a survey of blockchain and IAM startups, foundations, and initiatives but misses various key players such as Sovrin or Hyperledger Indy. Furthermore, the paper does not describe the purpose and goal of SSI or its building blocks.

Yang et al. [181] provide a summary of privacy-preserving technologies for blockchain. These technologies include ZKP, Ring Signatures, homomorphic cryptography, and other cryptographic algorithms. This survey does not focus on SSI or its building blocks or provide any detail on existing or emerging identity management models.

In [182], Lim et al. provide a definition of various identity models, in addition to explaining and comparing three popular blockchain implementations. Moreover, their paper states and compares several blockchain-based identity management projects including Sovrin, ShoCard, and uPort. However, their work lacks an in-depth discussion on the standards and privacy aspects of the SSI ecosystem as well as its building blocks and principles.

Paul et al. [183] assess the feasibility of deploying decentralized identity models through various important requirements. These requirements include secure delegation of credentials, user experience, and methods of interacting with the blockchain. Their work also discusses the questions around the optional approaches to utilizing the unique qualities of blockchain such as immutability and transparency, for identity management purposes. Nevertheless, their work lacks an in-depth overview of identity and identity management, privacy engineering techniques, or the legal and privacy aspect of the decentralized identity model.

In [184], Alexander provides an explanation of the SSI architecture and the essential components of SSI such as verifiable credentials. Furthermore, their work discusses the concept of identification and authentication in the context of SSI. However, their work does not discuss common SSI implementations and identity management models in detail, nor provides references to existing or emerging privacy engineering protocols.

In [159], Kuperberg provides a summary of the SSI model from the enterprise and ecosystem perspective and compares and contrasts various SSI implementations using 75 evaluation criteria. Their work analyzes existing implementation based on system maturity and the capabilities required to compete with conventional blockchain-less solutions. While their work provides a good summary of SSI, it misses a detailed explanation of the SSI history, principles, and building blocks, in addition to any discussion around privacy engineering techniques.

Finally, NIST has published a taxonomic survey of blockchain-based identity management systems [185]. Their work focuses on the terminology, model definition, and building blocks of SSI technology in addition to its privacy and security considerations. Their work, while thorough, is geared more toward the operational requirements rather than research efforts.

In comparison to the above surveys, this article fills in the missing gaps in the previous works. It provides a comprehensive overview of the digital identity and SSI, including its technical, architectural, and legal aspects as well as popular privacy engineering protocols that supplement SSI, and recent research advances in the field. As such, this article complements and enhances the reviews in previously published surveys.

*8.2. Original Research and Proposals.* Wang and De Filippi provide a description of the building blocks of SSI and discuss the Kiva case study which relies on DID, VC, and Hyperledger Indy [24]. The initial purpose of the Kiva protocol is to address the problems facing the microlending industry, whereby many constituents are ineligible for loans due to lack of formal identity and history. In their paper, they describe the importance of flexibility in any identity model to support the malleable nature of the human identity.

Shuaib et al. propose the use of SSI in the healthcare space and provide a brief summary of available scholarly work related to the healthcare use case [186]. In addition, they state the high-level requirements of SSI adoption in healthcare and summarize the advantages of using SSI in healthcare. In paper [187], Shuaib et al. discuss the use of the SSI model for the management of COVID-19 vaccination and test certifications.

Houtan et al. provide a survey on the use of blockchain and self-sovereign identity model for management of electric patient and health records [188]. The a uthors p rovide a classification of existing frameworks, proof of concepts, and published research work.

In paper [189], Naik and Jenkins evaluate the compatibility of uPort and Sovrin SSI ecosystem against GDPR principles, with the conclusion that those SSI systems are based on public permissioned blockchain are able to comply with the majority of key GDPR principles.

Kondova and Erbguth provide an assessment of Sovrin, uPort, and Jolocom SSI platforms and analyze the arising issues related to GDPR and SSI [190]. Their p aper raises various privacy concerns around SSI and focuses on the issues around the storage of four types of data, namely, DIDs, verifiable c redentials, r evocation d ata, a nd cryptographically hashed data.

Fedrecheski et al. describe the benefits and challenges of using DIDs and VCs to identify and authenticate Internet of Things (IoT) devices and their users [191]. The authors state the main challenges to be using constrained devices to perform asymmetric cryptography, heavy communication and DID resolution, and the privacy concerns around global tracking of devices, and finally lack of options in lightweight SSI software for IoT devices.

Belchior et al. propose an SSI-based access control (SSIBAC) to address the authorization and access control requirements of SSI [192]. Their p roposed m odel r elies on attribute-based model as well as selective disclosure of attributes, and range proofs for numerical values to ensure only the necessary data are disclosed to the authorized entities.

Liu et al. [193] compare traditional identity management systems with decentralized identity systems. Their assessment finds the differences stemming from their trust model and data storage schemes. They further identify maintaining privacy requirements and implementation of certain identity operations such as deletion, to be more difficult in a decentralized identity system.

In [194], Lux et al. provide a proof-of-concept integration of SSI and OpenID Connect protocol on Hyperledger Indy, with ability for the users to choose from a pool of identity providers instead of a select few. Their proposed system is able to receive OIDC claims as verifiable credentials.

Mukta et al. [195] propose CredChain, an SSI system that enables the creation, sharing, and verification of credentials, in addition to flexible and selective disclosure of claims using redactable signatures [196].

Soltani et al. [197] developed a financial client onboarding prototype using Hyperledger Indy to demonstrate the potential of SSI in streamlining client registration. This framework is evaluated against SSI and Privacy by Design principles, as well as relevant GDPR clauses concerning personal data management.

Othman and Callahan propose the Horcrux protocol [198], a decentralized authentication model, in which they rely on decentralized identifiers and IEEE 2410-2017 Biometric Open Protocol Standard (BOPS) to develop an SSI solution capable of storing decentralized biometric credentials.

Takemiya et al. propose the Sora SSI Identity Protocol [199]. This protocol relies on Hyperledger Iroha and DIDs and utilizes a central server to store encrypted copies of users' cryptographic keys and personal data. The Sora protocol supports the storage of salted hash representation of the users' identity data, along with the associated digital signatures generated by the identity issuers, on to the blockchain. This feature provides support for identity validation, nonrepudiation, and time-locking capabilities.

Dima et al. introduce the Scholarium platform [200] in which they rely on the MultiChain, a permissioned blockchain technology, to allow educational institutes to issue digital identity claims, such as graduation titles, to students. The Blockcert project [201], originally developed by the MIT Media Lab, is another pioneering blockchain-based solution for academic solutions.

Schanzenbach et al. developed reclaimID, an SSI platform for the peer-to-peer exchange of claim by utilizing the GNU Name System (GNS) [202]. The GNS system relies on a distributed hash table (DHT) and supports encrypted records, queries, and responses. The attributes stored on the name system are managed using attribute-based encryption (ABE) allowing users to selectively disclose required attributes with the requesting parties.

Finally, Abraham et al. propose a proof of concept to realize the derivation of digital identity attributes issued by traditional identity management systems into an SSI system based on Hyperledger Indy [203]. Their work also extends Hyperledger Indy's consensus protocol to support additional features including multisignature operations and verification of signatures contained in identity assertions.

SSI implementations are growing among various sectors including the financial services, healthcare, and government sector.

Various initiatives have been developed to address the challenges around knowing your customer (KYC) and anti-money laundering (AML) requirements. Know your customer describes the policies around the activities related to verifying the identity of a client and understanding and mitigating the risks associated with them.

Biryukov et al. proposed a privacy-preserving KYC scheme called KYCE based on the Ethereum blockchain [204]. The KYCE scheme allows users to take advantage of a unified identity to gain access to multiple financial services while preserving the privacy of the users by storing sensitive personal information with the KYC provider. Their work relies on ZKP cryptographic accumulator based on Jan et al. [205].

Moyano and Ross developed a KYC process based on DLT, in which a regulator maintains the database of users'

personal information [206]. Upon the first interaction between a user and a bank, the user's information is shared with and verified by the bank, and upon approval, the bank places the hash representation of the user's data in blockchain. Subsequent banks rely on the hash of the data to review and confirm the authenticity of the user's data.

KYCStart [207], KYC-Chain [208], and Cambridge Blockchain [209] are among the many projects that rely on DLT to deliver streamlined and privacy-preserving KYC schemes.

Similarly, in the health sector, patients' information is fragmented and out of the control of the patients. Various projects including MintHealth [210] have paved the way for infusing the SSI model into the current healthcare digital identity and trust frameworks. X. Liang et al. propose a mobile healthcare application that relies on a Hyperledger Fabric registering the hash of medical data, and cloud providers for secure storage of medical data where users have wide-ranging control over the management of their personal data [211].

Governments around the world are exploring the potential of SSI. The Canadian province of British Columbia has developed a platform called Verified Organization Network (VON) based on Hyperledger Indy, to allow organizations to manage their verified claims [212]. Likewise, the US's Department of Homeland Security's Science and Technology Directorate has invested in the development of SSI standards [213]. Estonia's e-Residency system provides applicants outside of Estonia the possibility to register a company or open a bank account [214]. Finally, Australia's post office and Digital Transformation Office have explored blockchain-based identity management systems [215].

We conclude this survey research by stating the future work ahead and the concluding remarks.

## 9. Conclusion and Future Research Directions

The field of self-sovereign identity is young and evolving. On one hand, there is rapid growth and commercialization of SSI platforms, and on the other hand, there is growing awareness and research on cryptographic protocols, and schemes empowering user's privacy.

Observing the current state of SSI, we have identified the following as the future research direction.

First, the user experience is a major avenue of research. Although the collective understanding of user experience has improved in the security space, only a small share of that change relates to key management. Losing a private key or password can lead to major vulnerabilities and economic costs. In many cases, when a secret key is lost, regenerating a new key and digital identity may not be feasible or serve little purpose. Proper key management in the context of SSI is a major factor in its mass adoption rate. In addition to key management, there are open research questions around the acceptable approaches by which users should be informed about, and consent to, sharing their data with other parties.

Secondly, similar to any new business initiative, the incentives and the return on investment on deployment, operation, and participating in an SSI network must be discussed and analyzed.

Another research path is the understanding of the necessary efforts around designing interoperable and consistent policies and specifications for SSI. For example, all entities engaged in an SSI network must conform to a set of standards to generate, ingest, display, exchange, and validate identity credentials successfully.

Moreover, it is important to realize and articulate the correct degree of decentralization that can support the vision and requirements of a user-centric identity model. More research must be done to better design critical identity operations such as identity issuance, user authentication, identity lookup, and secure data storage, as these actions may rely on some degree of centralization. Efforts to determine the correct balance between centralization and decentralization is an important research topic.

Design of fair governance and the trust frameworks suitable for SSI is another research opportunity. On one end of the spectrum, solutions based on Decentralized Autonomous Organization (DAO) that rely purely on smart contracts to provide completely decentralized governance policies have suffered from various vulnerabilities in the past. On the other end, models that rely on an agreed-upon trust framework which every member should accept and abide by before joining the network may be at the risk of unexpected and gradual centralization.

Finally, the growth and success of many SSI systems are entangled with addressing the challenges of underlying blockchain systems. Knowing that some SSI implementations require rapid interactions with blockchain and considering that a single interaction between two entities may involve multiple blockchain interactions, it is important to thoroughly evaluate the scalability, operational cost, and performance of the underlying blockchain system.

The purpose of this survey is to provide a comprehensive overview of the self-sovereign identity model and highlight its building blocks, well-known implementations, opportunities, and challenges. Identity is a central pillar of trust, and identity and access management is a vibrant, multidisciplinary, and growing field that requires attention, research, experimentation, and collaboration. SSI is still in its infancy and is not a silver bullet. However, in the age of surveillance capitalism, SSI is a step towards a more user-centric identity model with many potentials.

## Appendix

Claim: a statement about an entity. A claim can be made by the subject itself or by another entity. See Section 3.3.

Decentralized identifier (DID): a globally unique and persistent identifier that does not need a centralized registration authority because it is generated cryptographically. See Section 4.2.

Decentralized identity management: an identity management model that extends the authority for identifier generation and management beyond the traditional central root of trust. See Section 4.

Decentralized public key infrastructure (DPKI): a public key infrastructure that does not rely on the traditional certificate authorities for certificate generation but rather uses protocols such as DID to discover and verify public key descriptions. See Section 4.4.

Digital identity: a digital identity is a representation of an entity within a particular digital context. See Section 3.2.

Distributed ledger technology (DLT): a consensus of replicated, shared, and synchronized digital data geographically spread across multiple locations. See Section 4.1.

Identifier: a value used as a reference to a real-world identity or a specific persona. See Sections 3.3 and 4.2.

Identity: the set of behavioral or personal characteristics by which an entity is recognizable. Identity may apply to any type of entity, including individuals, organizations, and things. See Section 3.1.

Identity and access management (IAM): a collection of tools, processes, and policies used to manage individual identities, their authentication, authorization, roles, and privileges, within an organization or across boundaries. See Section 3.3.

Self-sovereign identity (SSI): an identity management model based on the main principle that the identity holder can control one or more identifiers and their identity data. See Sections 3.5.1 and 4.

## Data Availability

The underlying data can be found from various research repositories including Hindawi, IEEE, ACM, Google Scholar, and others.

## Conflicts of Interest

The authors declare that they have no conflicts of interest.

## References

[1] "1.1 billion invisible people without id are priority for new high level advisory council on identification for development," 2017, https://www.worldbank.org/en/news/press-release/2017/10/12/11-billioninvisible-people-without-id-are-priority-for-new-highlevel-advisory-council-on-identification-for-development.

[2] C. Satchell, G. Shanks, S. Howard, and J. Murphy, "Identity crisis: user perspectives on multiplicity and control in federated identity management," *Behaviour & Information Technology*, vol. 30, no. 1, pp. 51–62, 2011.

[3] "Case C-311/18, Data Protection Commissioner V. Facebook Ireland Limited and Maximillian Schrems," 2020, https://eur-lex.europa.eu/legal-content/EN/TXT/?uri=%20CELEX%3A62018CJ0311&qid=1616638710605 (visited on 03/26/2021).

[4] A. Mitchell and J. Smith, "Economics of Identity: The Size and Potential of the UK Market for Identity Assurance," 2015.

[5] General data protection regulation (GDPR), "*art. 4 GDPR*. General data protection regulation (GDPR)," 2016, https://gdpr-info.eu/%20art-4-gdpr/%20(visited%20on%2002/21/2021).

[6] C. Gallo, A. Jumamil, and P. Aranyawat, "Blockchain and Financial Inclusion: The Role Blockchain Technology Can Play in Accelerating Financial Inclusion," 2017.

[7] S. Zuboff, *The Age of Surveillance Capitalism: The Fight for a Human Future at the New Frontier of Power: Barack Obama's Books of 2019*, Profile books, London, UK, 2019.

[8] "Android User Locked Out of Google Accounts after Moving to A New City - Slashdot," 2016, https://tech.slashdot.org/story/16/11/21/%200032213/android-user-locked-out-of-google-accounts-aftermoving-to-a-new-city.

[9] "LastPass reveals 8 truths about passwords in the new password exposé - the LastPass blog," 2017, https://blog.lastpass.com/2017/%2011/lastpass-reveals-8-truths-about-passwords-in-the-newpassword-expose.html/.

[10] Data Breaches Compromised 3, "3 Billion Records in First Half of 2018*," 2021, https://www.gemalto.com/press/pages/data-breachescompromised-3-3-billion-records-in-first-half-of-2018.aspx.

[11] "Identity Fraud: Fraud Enters a New Era of Complexity — Javelin," 2018, https://www.javelinstrategy.com/coverage-area/2018identity-fraud-fraud-enters-new-era-complexity.

[12] "Equifax breach exposed millions of driver's licenses, phone numbers, emails — ars Technica," 2018, https://arstechnica.com/informationtechnology/2018/05/equifax-breach-exposed-millions-ofdrivers-licenses-phone-numbers-emails/.

[13] G. A. Akerlof and R. E. Kranton, "Identity economics: how our identities shape our work, wages, and well-being," in *Economics Books*Princeton University Press, NJ, USA, 2011.

[14] "Definition of IDENTITY," 2021, https://www.merriam-webster.com/dictionary/identity.

[15] Houghton Mifflin Harcourt Publishing Company, "The American Heritage Dictionary entry: identity," 2021, https://www.ahdictionary.com/%20word/search.html?q=identity.

[16] G. B. Ayed, *Architecting User-Centric Privacy-As-A-Set-Of-Services: Digital Identity-Related Privacy Framework*, Springer, NY, USA, 2014.

[17] D. F. James, *What is identity (as we now use the word)*, Stanford University, CA, USA, 1999.

[18] R. Paul, *Oneself as Another*, University of Chicago Press, Chicago, USA, 1992.

[19] M. S. Ferdous et al., "Security usability of petname systems," in *In Proceedings of the Nordic Conference on Secure IT Systems*, pp. 44–59, Springer, Oslo, Norway, October 2009.

[20] N. Strohminger, J. Knobe, and G. Newman, "The true self: a psychological concept distinct from the self," *Perspectives on Psychological Science*, vol. 12, pp. 551–560, 2017.

[21] E. C. . James, "Sociological perspectives on identity formation: the culture–identity link and identity capital," *Journal Of Adolescenc*, vol. 19, pp. 417–428, 1996.

[22] H. Erikson, *Childhood and Society*, WW Norton & Company, NY, USA, 1993.

[23] U. Glässer and M. Vajihollahi, "Identity management architecture," in *Security Informatics*, pp. 97–116, Springer, NY, USA, 2010.

[24] F. Wang and P. D Filippi, "Self-sovereign identity in a globalized world: credentials-based identity systems as a

driver for economic inclusion," *Frontiers in Blockchain*, vol. 2, 2020.
[25] D. Birch, *Identity Is the New Money (Perspectives)*, London Publishing Partnership, London, UK, 2014.
[26] G. Paul, M. Garcia, and J. Fenton, "Digital Identity Guidelines (Revised Draft)," Technical Report, National Institute of Standards and Technology, MD, USA, 2017.
[27] T. Itu, "Telecommunication standardization sector of ITU," *Annex C RTP Payload Format H*, vol. 261, pp. 108–113, 1993.
[28] K. Cameron, "The laws of identity," *Micro*, vol. 5, pp. 8–11, 2005.
[29] S. Clauß and M. Köhntopp, "Identity management and its support of multilateral security," *Computer Networks*, vol. 37, no. 2, pp. 205–219, 2001.
[30] M. S. Ferdous, G. Norman, and P. Ron, "Mathematical modelling of identity, identity management and other related topics," in *Proceedings of the 7th International Conference on Security of Information and Networks. ACM*, Glasgow, UK, September 2014.
[31] M. S. Ferdous, F. Chowdhury, and M. O Alassafi, "In search of self-sovereign identity leveraging blockchain technology," *IEEE Access*, vol. 7, Article ID 103059, 2019.
[32] F. G. Mármol, J. Girao, and G. M. Pérez, "TRIMS, a privacy-aware trust and reputation model for identity management systems," *Computer Networks*, vol. 54, no. 16, pp. 2899–2912, 2010.
[33] "Identity and Access Management-Global Market Outlook (2016-2022) — Orbis Research," 2017, https://www.orbisresearch.com/reports/%20index/identity-and-access-management-global-market-outlook2016-2022.
[34] Nobody Knows You're a Dog: https://www.washingtonpost.com/%20blogs/comic-riffs/post/nobody-knows-youre-a-dog-as-iconicinternet-cartoon-turns-20-creator-peter-steiner-knows-thejoke-rings-as-relevant-as-ever/2013/07/31/73372600-f98d11e2-8e84-c56731a202fb_blog.html.
[35] E. McCallister, *Guide to Protecting the Confidentiality of Personally Identifiable Information*, Diane Publishing, PA, USA, 2010.
[36] "Verifiable Credentials Data Model 1.0," 2021, https://www.w3.org/TR/%20verifiable-claims-data-model/.
[37] "trust — Definition of trust in English by Oxford Dictionaries," http://https://en.oxforddictionaries.com/definition/trust%20(visited%20on%2006/17/2021).
[38] Ping Identity, "Internet-Scale Identity Systems: An Overview and Comparison,“ 2017.
[39] Z. Wilcox-O'Hearn, "Names: Decentralized, Secure, Human-Meaningful: Choose Two," 2013.
[40] FIDO Alliance, "FIDO alliance," 2015.
[41] "IEEE 2410-2015 - IEEE Standard for Biometric Open Protocol," 2019, https://standards.ieee.org/standard/2410-2015.html.
[42] "Rebooting the web-of-trust," 2019, http://www.weboftrust.info/.
[43] "RWOT6 in Santa Barbara, California," 2018, https://github.com/WebOfTrustInfo/%20rwot6-santabarbara/blob/master/draft-documents/Biometrics.%20md.
[44] I. Riabi, H. K. B. Ayed, and L. A. Saidane, "A survey on Blockchain based access control for Internet of Things," in *Proceedings of the 2019 15th International Wireless Communications Mobile Computing Conference (IWCMC)*, pp. 502–507, Tangier, Morocco, June 2019.
[45] D. Rotondi and S. Piccione, "Managing access control for things: a capability based approach," *BodyNets*, pp. 263–268, 2012.
[46] T. Midhun, K. Prasanth, and J. Anoop, "A survey on authorization systems for web applications," *Journal of Computer Engineering (IOSR-JCE)*, vol. 17, no. 3, pp. 01–05, 2015.
[47] A. Jøsang and S. Pope, "User centric identity management," in *Proceedings of the AusCERT Asia Pacific Information Technology Security Conference*, p. 77, Brisbane, Australia, 2005.
[48] J. McWaters, *A Blueprint for Digital Identity*, 2016.
[49] B. C. Neuman and T. Ts'o, "Kerberos: an authentication service for computer networks," *IEEE Communications Magazine*, vol. 32, no. 9, pp. 33–38, 1994.
[50] Itu-T., "Recommendation database," https://www.itu.int/ITUT/recommendations/rec.aspx?rec=X.509.
[51] C. Allen, A. Brock, and V. Buterin, "Decentralized Public Key Infrastructure - A White Paper from Rebooting the Web of Trust," 2015, http://www.weboftrust.info/downloads/dpki.%20pdf.
[52] *Certinomis: Certificate for test.Com, O=Entreprise TEST*. url: https://bugzilla.mozilla.org/show_bug.cgi?id=1496088 (visited on 05/06/2021.
[53] N. V. D Meulen, "Diginotar: dissecting the first Dutch digital disaster," *Journal of Strategic Security*, vol. 6, no. 2, pp. 46–58, 2013.
[54] "How the Comodo Certificate Fraud Calls CA Trust into Question," 2021, https://arstechnica.com/information-technology/2011/03/howthe-comodo-certificate-fraud-calls-ca-trust-into-question/.
[55] "Trustwave admits issuing man-in-the-middle digital certificate," 2021, https://www.computerworld.com/article/%202501291/trustwave-admits-issuing-man-in-the-middle-digitalcertificate--mozilla-debates-punishment.html.
[56] D. W. Chadwick, "Federated identity management," in *Foundations of Security Analysis and Design V*, pp. 96–120, Springer, NY, USA, 2009.
[57] Saml Specifications — Saml X. M. L.: 2021, http://saml.xml.org/ saml-specifications.
[58] E. H. Lahav, "The OAuth 1.0 Protocol," https://tools.ietf.org/html/rfc5849.
[59] "Welcome to OpenID Connect – OpenID," https://openid.net/connect/.
[60] H. L'Amrani, B. E. Berroukech, Y. E. B. E. Idrissi, and R. Ajhoun, "Identity management systems: laws of identity for models 7 evaluation," in *Proceedings of 2016 4th IEEE International Colloquium on Information Science and Technology (CiSt)*, pp. 736–740, IEEE, 2016.
[61] "New Tech: Decentralized Digital Identity (DDID), Q1 2020," http://https://www.forrester.com/report/New+Tech+Decentralized+Digital+%20Identity+DDID+Q1+2020/-/E-RES147115/.
[62] Blockchain, "The dawn of decentralized identity," 2016, http://gartner.com/en/documents/3464117/blockchain-the-dawn-of-decentralized-identity.
[63] "Decentralized ID and self-sovereign identity are in the works," http://https://www.mercatoradvisorygroup.com/Templates/BlogPost.aspx?id=%207834&blogid=25506/.
[64] R. C. Merkle, "Secure communications over insecure channels," *Communications of the ACM*, vol. 21, no. 4, pp. 294–299, 1978.
[65] "What is"sovereign source authority"?," 2012, https://www.moxytongue.com/2012/02/what-is-sovereign-source-authority.html.
[66] C. Allen, *The Path to Self-Sovereign Identity. Life with Alacrity*, 2016.

[67] A. Preukschat and D. Reed, *Self-Sovereign Identity: Decentralized Digital Identity and Verifiable Credential*, Manning Publications, NY, USA, 2021.

[68] R. Eliezerov, *The Digital Identity Crisis. How the Explosion of Personal Information Is Transforming Technology*, Business and Society. Wiley, NJ, USA, 2018.

[69] R. Soltani, "Uyan Trang nguyen, and Aijun an "decentralized and privacy-preserving key management model," in *Proceedings of the 2020 International Symposium on Networks, Computers and Communications (ISNCC)*, pp. 1–7, Montreal, Canada, October 2020.

[70] "General Data Protection Regulation (GDPR) – Official Legal Text," 2020, https://gdpr-info.eu/.

[71] A. Angelov, M. Milkov, and M. Sørensen, *Decentralized Identity Management System for Self-Sovereign Identity*, Master Thesis, Alborg University, Alborg, Denmark, 2018.

[72] U. W. Chohan, "The decentralized autonomous organization and governance issues," *SSRN Electronic Journal*, https://papers.ssrn.com/sol3/papers.cfm?abstract_id=3082055, 2017.

[73] S. Garfinkel, *PGP: Pretty Good Privacy*, O'Reilly Media, Inc., CA, USA, 1995.

[74] "W3C Verifiable Credentials Working Group," 2021, https://www.w3.org/2017/vc/WG/.

[75] "Decentralized Identifiers (DIDs) v0.11," 2021, https://www.w3.org/TR/did-core/.

[76] P. J. Leach, M. Mealling, and R. Salz, "A Universally Unique IDentifier (UUID) URN Namespace," 2021, https://datatracker.ietf.org/doc/html/rfc4122.

[77] Json-Ld 1, https://json-ld.org/spec/latest/json-ld/, 2021.

[78] "DID method registry," 2019, https://w3c-ccg.github.io/did-methodregistry/.

[79] "IPFS is the distributed web," 2021, https://ipfs.io/.

[80] "Veres One - A Globally Interoperable Blockchain for Identity," 2021, http://https://veres.one/.

[81] "Peer DID method specification," 2021, https://identity.foundation/%20peer-did-method-spec/.

[82] "Universal Resolver Implementation and Drivers. Contribute to Decentralizedidentity/universal-Resolver Development by Creating an Account on GitHub," 2019, https://github.com/decentralized-identity/%20universal-resolver.

[83] "DIDComm messaging," 2021, https://identity.foundation/didcommmessaging/spec/.

[84] "The JavaScript object notation (JSON) data interchange format,".

[85] "JSON Web Token (JWT)," 2021, https://datatracker.ietf.org/doc/%20html/rfc7519.

[86] D. Cooper, S. Santesson, S. Farrell, S. Boeyen, R. Housley, and W. Polk, Internet X. 509 public key infrastructure certificate and certificate revocation list (CRL) profile," *RFC 5280*, pp. 1–151, 2008, https://datatracker.ietf.org/doc/html/rfc5280.

[87] I. C. A. NN., https://www.icann.org/%20, 2021.

[88] R. Scott et al., "Why Johnny still, still can't encrypt: evaluating the usability of a modern PGP client," 2015, https://arxiv.org/abs/1510.08555.

[89] IANA Stewardship Transition Coordination Group (ICG), "Proposal to transition the stewardship of the internet assigned numbers authority (IANA) functions," Technical Report, https://www.icann.org/en/system/files/files/iana-stewardship-transition-proposal-10mar16-en.pdf, 2016.

[90] Internet Corporation for Assigned Names and Numbers (ICANN), "CCWG-Accountability Supplemental Final Proposal on Work Stream 1 Recommendations," Technical Report, 2016.

[91] "Stewardship of IANA Functions Transitions to Global Internet Community as Contract with U.S. Government Ends," 2021, https://www.icann.org/en/announcements/details/stewardship-of-iana-functions-transitions-to-global-internet-community-as-contract-with-us-government-ends-1-10-2016-en.

[92] DIF-Decentralized Identity Foundation, 2021, https://identity.foundation/.

[93] DIF Sidetree Protocol, 2021, https://identity.foundation/sidetree/spec/.

[94] P. Sivakumar and K. Singh, "Privacy based decentralized public key infrastructure (PKI) implementation using smart contract in blockchain," Technical Report, 2017.

[95] B. Laurie, A. Langley, and E. Kasper, "Certificate transparency,".

[96] W. Gao, W. G. Hatcher, and W. Yu, "A survey of blockchain: techniques, applications, and challenges," in *Proceedings of the 2018 27th International Conference on Computer Communication and Networks (ICCCN)*, pp. 1–11, Hangzhou, China, July 2018.

[97] "Hyperledger Fabric – Hyperledger," 2021, https://www.hyperledger. org/projects/fabric.

[98] "Hyperledger Indy – Hyperledger," 2021, https://www.hyperledger.org/%20projects/hyperledger-indy.

[99] S. Seibold and S. George, *Consensus, immutable agreement for the internet of value*, KPMG, Amstelveen, Netherlands, 2016.

[100] M. Castro and B. Liskov, "Practical Byzantine fault tolerance and proactive recovery," *ACM Transactions on Computer Systems (TOCS) 20*, vol. 4, pp. 398–461, 2002.

[101] L. M. Bach, B. Mihaljevic, and M. Zagar, "Comparative analysis of blockchain consensus algorithms," in *Proceedings of the 2018 41st International Convention on Information and Communication Technology, Electronics and Microelectronics (MIPRO)*, pp. 1545–1550, IEEE, Opatija, Croatia, May 2018.

[102] M. Du, X. Ma, Z. Zhe, X. Wang, and Q. Chen, "A review on consensus algorithm of blockchain," in *Proceedings of the 2017 IEEE International Conference on Systems, Man, and Cybernetics (SMC)*, pp. 2567–2572, IEEE, Banff, AB, Canada, October 2017.

[103] D. Reed, J. Law, and D. Hardman, *The Technical Foundations of Sovrin, A White Paper from the Sovrin Foundation*, Sovrin, MA, USA, 2016.

[104] Sovrin Board of Trustees, *Sovrin provisional trust framework*, Sovrin, MA, USA, 2017.

[105] "Contribute To Hyperledger/indy-Agent Development by Creating an Account on GitHub," 2019, http://https://github.com/hyperledger/indy-agent.

[106] "How to use Google Duplex to make a restaurant reservation - the Verge," 2021, https://www.theverge.com/2018/12/5/18123785/googleduplex-how-to-use-reservations.

[107] "Storage And Compute Nodes for Decentralized Identity Data and Interactions: Decentralized-Identity/identity-Hub," 2019.

[108] "Solid: your data, your choice. Advancing Web standards to empower people," 2021, https://solidproject.org/.

[109] S. Brooks, M. Garcia, N. Lefkovitz, S. Lightman, and E. Nadeau, *An Introduction to Privacy Engineering and Risk Management in Federal Systems*, US Department of Commerce, National Institute of Standards and Technology, MD, USA, 2017.

[110] M. Sabt, M. Achemlal, and A. Bouabdallah, "Trusted Execution Environment: What it is, and what it is Not," in *Proceedings of the 2015 IEEE Trustcom/BigDataSE/ISPA*, pp. 57–64, Helsinki, Finland, Aug 2015.

[111] R. L. Rivest, A. Shamir, and Y. Tauman, "How to leak a secret," in *Proceedings of the International Conference on the Theory and Application of Cryptology and Information Security*, pp. 552–565, Springer, Berlin, Heidelberg, November 2001.

[112] A. C.-C. Yao, "Protocols for secure computations," in *Proceedings of the 23rd Annual Symposium on Foundations of Computer Science (sfcs 1982)*, pp. 160–164, IL, USA, November 1982.

[113] O. Goldreich, S. Micali, and A. Wigderson, "How to play any mental game," in *Proceedings of the Nineteenth Annual ACM Symposium on Theory of Computing*, ACM, New York, NY, USA, pp. 218–229, January 1987.

[114] B. Chor and N. Gilboa, "Computationally private information retrieval," in *Proceedings of the twenty-ninth annual ACM symposium on Theory of computing*, CA, USA, May 1997.

[115] Y. Lindell, "Secure multiparty computation for privacy preserving data mining," in *Encyclopedia of Data Warehousing and Mining*, pp. 1005–1009, IGI Global, Hershey, PA, 2005.

[116] D. G. Nair, V. P. Binu, and G. S. Kumar, "An effective private data storage and retrieval system using secret sharing scheme based on secure multi-party computation," in *Proceedings of the 2014 International Conference on Data Science Engineering (ICDSE)*, pp. 210–214, Kochi, India, August 2014.

[117] C. Clifton, K. Muratr, J. Vaidya, X. Lin, and M. Y. Zhu, "Tools for privacy preserving distributed data mining," *ACM Sigkdd Explorations Newsletter*, vol. 4, pp. 28–34, 2002.

[118] D. K. Mishra, R. Sheikh, and B. Kumar, "Privacy-preserving k-secure sum protocol," *IJCSIS International Journal of Computer Science and Information Security*, vol. 6, 2009.

[119] F. A. N. Pathak and S. B. S. Pandey, "An efficient method for privacy preserving data mining in secure multiparty computation," in *Proceedings of the 2013 Nirma University International Conference on Engineering (NUiCONE)*, pp. 1–3, Ahmedabad, India, November 2013.

[120] B. Pinkas, "Cryptographic techniques for privacy-preserving data mining," *ACM Sigkdd Explorations Newsletter*, vol. 4, no. 2, pp. 12–19, 2002.

[121] "Enigma. MIT Media Lab," 2021, https://www.media.mit.edu/projects/%20enigma/overview/.

[122] R. Soltani, U. T. Nguyen, and A. An, "Practical key recovery model for self-sovereign identity based digital wallets," in *2019 Proceedings of the IEEE Intl Conf on Dependable, Autonomic and Secure Computing, Intl Conf on Pervasive Intelligence and Computing, Intl Conf on Cloud and Big Data Computing, Intl Conf on Cyber Science and Technology Congress (DASC/PiCom/CBDCom/CyberSciTech)*, pp. 320–325, IEEE, Fukuoka, Japan, August 2019.

[123] A. Shamir, "How to share a secret," *Communications of the ACM*, vol. 22, no. 11, pp. 612-613, 1979.

[124] G. R. Blakley, "Safeguarding cryptographic keys," in *Proceedings of the 1979 International Workshop on Managing Requirements Knowledge (MARK)*, NY, USA, June 1979.

[125] Y. Zeng and D. Liu, "A key escrow scheme to IOT based on Shamir," in *Proceedingds of the 2013 International Conference on Communications, Circuits and Systems (ICCCAS)*, pp. 94–97, IEEE, Chengdu, China, November 2013.

[126] S. Goldwasser, S. Micali, and C. Rackoff, "The knowledge complexity of interactive proof systems," *SIAM Journal on Computing*, vol. 18, no. 1, pp. 186–208, 1989.

[127] J. Kilian, "A note on efficient zero-knowledge proofs and arguments," in *Proceedings of the Twenty-Fourth Annual ACM Symposium on Theory of Computing*, pp. 723–732, ACM, Victoria, British Columbia, Canada, May 1992.

[128] P. Bryan, H. Jon, G. Craig, and R. Mariana, "Pinocchio: nearly practical verifiable computation," in *Proceedings of the 2013 IEEE Symposium on Security and Privacy*, pp. 238–252, IEEE, Berkeley, CA, USA, May 2013.

[129] J. Groth, "On the size of pairing-based non-interactive arguments," in *Proceedings of the Annual International Conference on the Theory and Applications of Cryptographic Techniques*, pp. 305–326, Springer, Berlin, Heidelberg, April 2016.

[130] B. Bünz, J. Bootle, D. Boneh, A. Poelstra, P. Wuille, and G. Maxwell, "Bulletproofs: short proofs for confidential transactions and more," in *Proceedings of the 2018 IEEE Symposium on Security and Privacy (SP)*, pp. 315–334, IEEE, San Francisco, CA, USA, May 2018.

[131] C. Jan and L. Anna, "A signature scheme with efficient protocols," in *Proceedings of the International Conference on Security in Communication Networks*, pp. 268–289, Springer, Berlin, Heidelberg, March 2002.

[132] A. Fiat and A. Shamir, "How to prove yourself: practical solutions to identification and signature problems," in *Proceedings of the Conference on the Theory and Application of Cryptographic Techniques*, pp. 186–194, Springer, Berlin, Heidelberg, 1986.

[133] E. B. Sasson, A. Chiesa, D. Genkin, E. Tromer, and M. Virza, "SNARKs for C: verifying program executions succinctly and in zero knowledge," in *Proceedings of the Annual Cryptology Conference*, pp. 90–108, Springer, CA,USA, August 2013.

[134] E. B. Sasson, I. Bentov, Y. Horesh, and M. Riabzev, "Scalable, transparent, and post-quantum secure computational integrity," *In IACR Cryptology ePrint Archive 2018*, vol. 46, 2018.

[135] "Identity mixer, IBM research zurich," 2015, https://www.zurich.ibm.com/pdf/csc/Identity_Mixer_Nov_2015.pdf.

[136] U-Prove, https://www.microsoft.com/enus/research/project/u-prove/, 2021.

[137] M. Blum, P. Feldman, and S. Micali, "Non-interactive ZeroKnowledge and its applications," in *Proceedings Of the Twentieth Annual ACM Symposium On Theory Of Computing. STOC '88*, Association for Computing Machinery, Chicago, Illinois, USA, pp. 103–112, January 1988.

[138] H. Wu and F. Wang, "A survey of noninteractive zero knowledge proof system and its applications," *The Scientific World Journal*, vol. 2014, Article ID 560484, 7 pages, 2014.

[139] J. J. Quisquater, M. Quisquater, and M. Quisquater, "How to explain zero-knowledge protocols to your children," in *Proceedings pf the Conference on the Theory and Application of Cryptology*, pp. 628–631, Springer, Belgium, China, April 1989.

[140] T. Koen, C. Ramaekers, and V. W. Cees, *Efficient ZeroKnowledge Range Proofs in Ethereum*, ING, Amsterdam, Netherlands, 2017.

[141] "Evernym — The Self-Sovereign Identity Company," 2021, http://https://www.evernym.com/.

[142] Privacy-Enhancing Technology For Your Enterprise Blockchain — QEDIT, "A data privacy layer for your enterprise blockchain — qedit," 2021, https://qed-it.com/.

[143] ZKProof Standards, http://zkproof.org, 2021.

[144] "Engineering privacy for verified credentials," 2020, https://w3c-ccg.github.io/data-minimization/.

[145] R. L. Rivest, L. Adleman, M. L. Dertouzos et al., "On data banks and privacy homomorphisms," *Foundations of secure computation*, vol. 4, no. 11, pp. 169–180, 1978.

[146] C. Dwork, "Differential privacy: a survey of results," in *Proceedings of the International Conference on Theory and Applications of Models of Computation*, pp. 1–19, Springer, Berlin, Heidelberg, April 2008.

[147] A. Shamir, "Identity-based cryptosystems and signature schemes," in *Proceedings of the Workshop on the Theory and Application of Cryptographic Techniques*, pp. 47–53, Springer, Paris, France, April 1984.

[148] G. Brassard, D. Chaum, and C. Claude, "Minimum disclosure proofs of knowledge," *Journal of Computer and System Sciences*, vol. 37, no. 2, pp. 156–189, 1988.

[149] J. B. Bernabe, J. L. Canovas, J. H. L. Ramos, R. T. Moreno, and S. Antonio, "Privacy-preserving solutions for blockchain: review and challenges," *IEEE Access*, vol. 7, Article ID 164908, 2019.

[150] M. Ali, R. Shea, J. Nelson, and M. J. Freedman, "Blockstack: a new internet for decentralized applications," *Whitepaper*, 2017.

[151] "Civic Secure Identity Ecosystem - Decentralized Identity & Reusable KYC. En-US," 2021, https://www.civic.com/.

[152] uPort.me, 2021, https://www.uport.me/.

[153] "Hyperledger – Open Source Blockchain Technologies," 2021, https://www.hyperledger.org/.

[154] "Sovrin," 2021, https://sovrin.org/.

[155] "Jolocom," 2021, https://jolocom.io/.

[156] "Selfkey.," 2021, https://selfkey.org/.

[157] "lifeID," 2021, https://lifeid.io/%20(visited%20on%2006/17/2021.

[158] "Mattr," 2021, https://mattr.global/.

[159] M. Kuperberg, "Blockchain-based identity management: a survey from the enterprise and ecosystem perspective," *IEEE Transactions on Engineering Management*, vol. 67, no. 4, 2019.

[160] P. Dunphy and F. A. P. Petitcolas, "A first look at identity management schemes on the blockchain," *IEEE Security & Privacy*, vol. 16, pp. 20–29, 2018.

[161] A. Abraham, "Self-sovereign identity," *Styria. EGIZ. GV. A. T.*, 2017.

[162] "World Wide Web Consortium (W3C)," 2021, https://www.w3.org/.

[163] "W3C DID Working Group," 2021, https://www.w3.org/2019/did-wg/.

[164] "W3C Credentials Community Group. en-US," 2021.

[165] "Identity Working Group - Identity Working Group - Hyperledger Confluence," 2021, https://wiki.hyperledger.org/display/IWG/Identity+%20Working+Group.

[166] "IIW - internet identity workshop," 2021, https://internetidentityworkshop.com/.

[167] Digital I. D. & Authentication Council of Canada, "Digital ID & Authentication Council of Canada," 2021, https://diacc.ca/.

[168] "About – Kantara Initiative," 2019, https://kantarainitiative.org/%20about/.

[169] D. Weeraman, "User Managed Access — UMA 2.0," 2017, https://medium.com/@dewni.matheesha/user-managed-accessuma-2-0-bcecb1d535b3.

[170] "ID2020. https://id2020.org/.

[171] "One World Identity – The Nexus of the Identity Industry," https://oneworldidentity.com/.

[172] "Self-Sovereign Identity For Everyone!," 2021, https://ssimeetup.org/.

[173] "Trust Over IP," 2020, https://trustoverip.org/.

[174] D. J. Solove, *A brief history of information privacy law*SSRN, NY, USA, 2016.

[175] K. Todo and K. Sato, "Directive 2000/60/EC of the European Parliament and of the Council of 23 October 2000 establishing a framework for Community action in the field of water policy," pp. 66–106, Cambridge University Press, Cambridge, UK, 2002.

[176] "The seven foundational principles," 2021, https://www.ryerson.ca/pbdce/about/.

[177] E. U Regulation ("No 910/2014 of the European Parliament and of the Council of 23 July 2014 on electronic identification and trust services for electronic transactions in the internal market and repealing Directive 1999/93/EC," 2014, http://data.europa.eu/eli/reg/%202014/910/oj/eng.

[178] "Assembly Bill No. 375. https://leginfo.legislature.ca.gov/%20faces/billTextClient.xhtml?bill_id=201720180AB375.

[179] A. Grüner, M. Alexander, and C. Meinel, "On the relevance of blockchain in identity management," 2018, https://arxiv.org/abs/1807.08136.

[180] O. Jacobovitz, "Blockchain for identity management," Technical Report, The Lynne and William Frankel Center for Computer Science Department of Computer Science. Ben-Gurion University, Beer Sheva, Isreal, 2016.

[181] D. Yang, J. Gavigan, and Z. W. O'Hearn, "Survey of confidentiality and privacy preserving technologies for blockchains," R3 Confidential and Privacy Report, R3 Research, Gujarat, India, 2016.

[182] S. Y. Lim et al., "Blockchain technology the identity management and authentication service disruptor: a survey," *International Journal on Advanced Science, Engineering and Information Technology*, vol. 8, no. 4-2), 2018.

[183] D. Paul, L. Garratt, and F. Petitcolas, "Decentralizing digital identity: open challenges for distributed ledgers," in *Proceedings of the 2018 IEEE European Symposium on Security and Privacy Workshops (EuroS&PW)*, pp. 75–78, IEEE, London, UK, April 2018.

[184] M. Alexander, "A survey on essential components of a selfsovereign identity," *Computer Science Review*, vol. 30, pp. 80–86, 2018.

[185] L. Lesavre, P. Varin, P. Mell, M. Davidson, and J. Shook, "A taxonomic approach to understanding emerging blockchain identity management systems," 2019, https://arxiv.org/abs/1908.00929.

[186] M. Shuaib, S. Alam, M. S. Alam, and M. S. Nazir, "Self-sovereign identity for healthcare using blockchain," *Materials Today: Proceedings*, vol. 46, 2021, In press.

[187] M. Shuaib, S. Alam, M. S. Alam, and M. S. Nazir, "Immunity credentials using self-sovereign identity for combating COVID-19 pandemic," *Materials Today: Proceedings*, 2021.

[188] B. Houtan, A. S. Hafid, and D. Makrakis, "A survey on blockchain-based self-sovereign patient identity in healthcare," *IEEE Access*, vol. 8, Article ID 90494, 2020.

[189] N. Naik and P. Jenkins, "Your identity is yours: take back control of your identity using GDPR compatible self-sovereign identity," in *Proceedings of the 2020 7th International Conference on Behavioural and Social Computing (BESC)*, pp. 1–6, IEEE, Bournemouth UK, Novenmber 2020.

[190] G. Kondova and J. Erbguth, "Self-sovereign identity on public blockchains and the GDPR," in *Proceedings of the 35th Annual ACM Symposium on Applied Computing*, pp. 342–345, NY, USA, April 2020.

[191] G. Fedrecheski, J. M. Rabaey, L. C. P. Costa, P. C. C. Ccori, W. T. Pereira, and M. K. Zuffo, "Self-sovereign identity for IoT environments: a perspective," in *Proceedings of the 2020 Global Internet of Things Summit (GIoTS)*, pp. 1–6, IEEE, Dublin, Ireland, June 2020.

[192] R. Belchior, B. Putz, G. Pernul, M. Correia, A. Vasconcelos, and S. Guerreiro, "SSIBAC: self-sovereign identity based access control," in *Proceedings of the 2020 IEEE 19th International Conference on Trust, Security and Privacy in Computing and Communications*, pp. 1935–1943, TrustCom), Guangzhou, China, December 2020.

[193] J. Liu, A. Hodges, L. Clay, and J. Monarch, "An analysis of digital identity management systems - a two-mapping view," in *Proceedings of the 2020 2nd Conference on Blockchain Research Applications for Innovative Networks and Services (BRAINS)*, pp. 92–96, Paris, France, September 2020.

[194] Z. A. Lux, D. Thatmann, S. Zickau, and F. Beierle, "Distributed-Ledger-based authentication with decentralized identifiers and verifiable credentials," in *Proceedings of the 2020 2nd Conference on Blockchain Research Applications for Innovative Networks and Services (BRAINS)*, pp. 71–78, Paris, France, September 2020.

[195] R. Mukta, J. Martens, H. Y. Paik, Q. Lu, and S. S. Kanere, "Blockchain-based verifiable credential sharing with selective disclosure," in *Proceedings of the 2020 IEEE 19th International Conference on Trust, Security and Privacy in Computing and Communications (TrustCom)*, pp. 959–966, Guangzhou, China, December 2020.

[196] R. Johnson, M. David, S. David, and W. David, "Homomorphic signature schemes," in *Proceedings of the Cryptographers' Track at the RSA Conference*, pp. 244–262, Springer, Berlin, Heidelberg, February 2002.

[197] R. Soltani, U. T. Nguyen, and A. An, "A new approach to client onboarding using self-sovereign identity and distributed ledger," in *Proceedings of the 2018 IEEE International Conference on Internet of Things (iThings) and IEEE Green Computing and Communications (GreenCom) and IEEE Cyber, Physical and Social Computing (CPSCom) and IEEE Smart Data (SmartData)*, pp. 1129–1136, IEEE, Halifax, NS, Canada, July 2018.

[198] A. Othman and J. Callahan, "The Horcrux protocol: a method for decentralized biometric-based self-sovereign identity," in *Proceedings of the 2018 International Joint Conference on Neural Networks (IJCNN)*, pp. 1–7, Rio de Janeiro, Brazil, July 2018.

[199] M. Takemiya and B. Vanieiev, "Sora identity: secure, digital identity on the blockchain," in *Proceedings of the 2018 IEEE 42nd Annual Computer Software and Applications Conference (COMPSAC)*, pp. 582–587, IEEE, Tokyo, Japan, July 2018.

[200] G. A. Dima, A. G. Jitariu, C. Pisa, and G. Bianchi, "Scholarium: supporting identity claims through a permissioned blockchain," in *Proceedings of the 2018 IEEE 4th International Forum on Research and Technology for Society and Industry (RTSI)*, pp. 1–6, Palermo, Italy, September 2018.

[201] P. Schmidt, "Blockcerts—an open infrastructure for academic credentials on the blockchain," 2016.

[202] S. Martin, G. Bramm, and J. Schütte, "Reclaimid: secure, self-sovereign identities using name systems and AttributeBased encryption," in *Proceedings of the 2018 17th IEEE International Conference on Trust, Security and Privacy in Computing and Communications/12th IEEE International Conference on Big Data Science and Engineering (TrustCom/BigDataSE)*, pp. 946–957, IEEE, NY, USA, August 2018.

[203] A. Abraham, T. Kevin, and E. Kirchengast, "Qualified eID derivation into a distributed ledger based IdM system," in *Proceedings of the 2018 17th IEEE International Conference on Trust, Security and Privacy in Computing and Communications/12th IEEE International Conference on Big Data Science and Engineering (TrustCom/BigDataSE)*, pp. 1406–1412, IEEE, NY, USA, August 2018.

[204] A. Biryukov, D. Khovratovich, and S. Tikhomirov, "Privacypreserving KYC on Ethereum," in *In Proceedings of the European Society for Socially Embedded Technologies (EUSSET)*, Nancy, France, June 2018.

[205] C. Jan, M. Kohlweiss, and C. Soriente, "An accumulator based on bilinear maps and efficient revocation for anonymous credentials," in *Proceedings of the International Workshop on Public Key Cryptography*, pp. 481–500, Springer, CA, USA, March 2009.

[206] J.P. Moyano and O. Ross, "KYC optimization using distributed ledger technology," *Business & Information Systems Engineering*, vol. 59, 2017.

[207] "Deloitte Digital - Blockchain Proof-Of-Concept to Mutualize KYC Checks — Deloitte Luxembourg — Press Release," 2020, https://www2.deloitte.com/lu/en/pages/technology/articles/%20deloitte-develops-blockchain-poc-kyc-checks.html.

[208] "Kyc-Chain - Blockchain and Banking KYC / AML Compliance Solution," 2020, https://kyc-chain.com/.

[209] "Cambridge blockchain — identity solutions," 2021, https://www.cambridgeblockchain.com/.

[210] MintHealth - Decentralized Health Information, "MintHealth - Decentralized Health Information," 2021, https://www.minthealth.io/.

[211] X. Liang, J. Zhao, S. Shetty, J. Liu, and D. Li, "Integrating blockchain for data sharing and collaboration in mobile healthcare applications," in *Proceedings of the 2017 IEEE 28th Annual International Symposium on Personal, Indoor, and Mobile Radio Communications (PIMRC)*, pp. 1–5, IEEE, Montreal, QC, Canada, October 2017.

[212] "Verifiable Organizations Network," 2021, https://vonx.io/.

[213] N. Release, "DHS S&T Awards $749K to Evernym for Decentralized Key Management Research and Development," Department of Homeland Security, 2017, https://www.dhs.gov/science-andtechnology/news/2017/07/20/news-release-dhs-st-awards-749kevernym-decentralized-key.

[214] e-Estonia, "We Have Built a Digital Society and We Can Show You How," 2020, https://e-estonia.com/.

[215] "DTO wants help designing whole-of-govt digital identities - strategy iTnews," 2020, https://www.itnews.com.au/news/dto-wants-helpdesigning-whole-of-govt-digital-identities-419101.